\documentclass[12pt]{article}
\usepackage{amsmath,amsthm,latexsym,amssymb,amsfonts,epsfig,psfrag,mathrsfs,bbm}

\newcommand{\fcmt}[2]{\fbox{\mbox{\begin{minipage}[t]{#1cm} #2 \end{minipage}}}}
\newcommand{\cmt}[2]{\mbox{\begin{minipage}[t]{#1cm} #2 \end{minipage}}}
\newcommand{\mb}[1]{\mathbbm{#1}}  
\newcommand{\Muserfunction}[1]{A}               

\makeatletter
\@addtoreset{equation}{section}
\makeatother

\def\be{\begin{equation}}
\def\ee{\end{equation}}
\def\ba{\begin{eqnarray}}
\def\ea{\end{eqnarray}}
\def\dprime{{\prime\prime}}
\def\Nl{{\mathchoice
{\setbox0=\hbox{$\displaystyle\rm N$}\hbox{\hbox to0pt
{\kern0.4\wd0\vrule height0.9\ht0\hss}\box0}}
{\setbox0=\hbox{$\textstyle\rm N$}\hbox{\hbox to0pt
{\kern0.4\wd0\vrule height0.9\ht0\hss}\box0}}
{\setbox0=\hbox{$\scriptstyle\rm N$}\hbox{\hbox to0pt
{\kern0.4\wd0\vrule height0.9\ht0\hss}\box0}}
{\setbox0=\hbox{$\scriptscriptstyle\rm N$}\hbox{\hbox to0pt
{\kern0.4\wd0\vrule height0.9\ht0\hss}\box0}}}}
\def\Zl{{\mathchoice
{\setbox0=\hbox{$\displaystyle\rm Z$}\hbox{\hbox to0pt
{\kern0.4\wd0\vrule height0.9\ht0\hss}\box0}}
{\setbox0=\hbox{$\textstyle\rm Z$}\hbox{\hbox to0pt
{\kern0.4\wd0\vrule height0.9\ht0\hss}\box0}}
{\setbox0=\hbox{$\scriptstyle\rm Z$}\hbox{\hbox to0pt
{\kern0.4\wd0\vrule height0.9\ht0\hss}\box0}}
{\setbox0=\hbox{$\scriptscriptstyle\rm Z$}\hbox{\hbox to0pt
{\kern0.4\wd0\vrule height0.9\ht0\hss}\box0}}}}
\def\Ql{{\mathchoice
{\setbox0=\hbox{$\displaystyle\rm Q$}\hbox{\hbox to0pt
{\kern0.4\wd0\vrule height0.9\ht0\hss}\box0}}
{\setbox0=\hbox{$\textstyle\rm Q$}\hbox{\hbox to0pt
{\kern0.4\wd0\vrule height0.9\ht0\hss}\box0}}
{\setbox0=\hbox{$\scriptstyle\rm Q$}\hbox{\hbox to0pt
{\kern0.4\wd0\vrule height0.9\ht0\hss}\box0}}
{\setbox0=\hbox{$\scriptscriptstyle\rm Q$}\hbox{\hbox to0pt
{\kern0.4\wd0\vrule height0.9\ht0\hss}\box0}}}}
\def\Rl{{\mathchoice
{\setbox0=\hbox{$\displaystyle\rm R$}\hbox{\hbox to0pt
{\kern0.4\wd0\vrule height0.9\ht0\hss}\box0}}
{\setbox0=\hbox{$\textstyle\rm R$}\hbox{\hbox to0pt
{\kern0.4\wd0\vrule height0.9\ht0\hss}\box0}}
{\setbox0=\hbox{$\scriptstyle\rm R$}\hbox{\hbox to0pt
{\kern0.4\wd0\vrule height0.9\ht0\hss}\box0}}
{\setbox0=\hbox{$\scriptscriptstyle\rm R$}\hbox{\hbox to0pt
{\kern0.4\wd0\vrule height0.9\ht0\hss}\box0}}}}
\def\Cl{{\mathchoice
{\setbox0=\hbox{$\displaystyle\rm C$}\hbox{\hbox to0pt
{\kern0.4\wd0\vrule height0.9\ht0\hss}\box0}}
{\setbox0=\hbox{$\textstyle\rm C$}\hbox{\hbox to0pt
{\kern0.4\wd0\vrule height0.9\ht0\hss}\box0}}
{\setbox0=\hbox{$\scriptstyle\rm C$}\hbox{\hbox to0pt
{\kern0.4\wd0\vrule height0.9\ht0\hss}\box0}}
{\setbox0=\hbox{$\scriptscriptstyle\rm C$}\hbox{\hbox to0pt
{\kern0.4\wd0\vrule height0.9\ht0\hss}\box0}}}}
\def\Hl{{\mathchoice
{\setbox0=\hbox{$\displaystyle\rm H$}\hbox{\hbox to0pt
{\kern0.4\wd0\vrule height0.9\ht0\hss}\box0}}
{\setbox0=\hbox{$\textstyle\rm H$}\hbox{\hbox to0pt
{\kern0.4\wd0\vrule height0.9\ht0\hss}\box0}}
{\setbox0=\hbox{$\scriptstyle\rm H$}\hbox{\hbox to0pt
{\kern0.4\wd0\vrule height0.9\ht0\hss}\box0}}
{\setbox0=\hbox{$\scriptscriptstyle\rm H$}\hbox{\hbox to0pt
{\kern0.4\wd0\vrule height0.9\ht0\hss}\box0}}}}
\def\Ol{{\mathchoice
{\setbox0=\hbox{$\displaystyle\rm O$}\hbox{\hbox to0pt
{\kern0.4\wd0\vrule height0.9\ht0\hss}\box0}}
{\setbox0=\hbox{$\textstyle\rm O$}\hbox{\hbox to0pt
{\kern0.4\wd0\vrule height0.9\ht0\hss}\box0}}
{\setbox0=\hbox{$\scriptstyle\rm O$}\hbox{\hbox to0pt
{\kern0.4\wd0\vrule height0.9\ht0\hss}\box0}}
{\setbox0=\hbox{$\scriptscriptstyle\rm O$}\hbox{\hbox to0pt
{\kern0.4\wd0\vrule height0.9\ht0\hss}\box0}}}}

\title{On (Cosmological) Singularity Avoidance in\\ Loop Quantum Gravity}
\author{
J. 
Brunnemann\thanks{jbrunnem@aei.mpg.de,jbrunnemann@perimeterinstitute.ca},
T. 
Thiemann\thanks{thiemann@aei.mpg.de,tthiemann@perimeterinstitute.ca}\\
\\
MPI f. Gravitationsphysik, Albert-Einstein-Institut, \\
           Am M\"uhlenberg 1, 14476 Potsdam, Germany\\
\\
and\\
\\
Perimeter Institute for Theoretical Physics,\\ 
31 Caroline Street N, Waterloo, ON N2L 2Y5, Canada}

\date{{\small Preprint AEI-2005-098}}

\begin{document}

\maketitle

\begin{abstract}
Loop Quantum Cosmology (LQC), mainly due to Bojowald, is not the 
cosmological sector of Loop 
Quantum Gravity (LQG). Rather, LQC consists of a truncation of the phase 
space of classical General Relativity to spatially homogeneous situations 
which is then quantized by the methods of LQG. Thus, LQC is a quantum 
mechanical toy model (finite number of degrees of freedom) for LQG  
(a genuine QFT with an infinite number of degrees of freedom) which
provides important consistency checks. However, it is 
a non trivial question whether the predictions of LQC are robust after 
switching on the inhomogeneous fluctuations present in full LQG.

Two of the most spectacular findings of LQC are that 1. 
the inverse scale factor is bounded from above on zero volume 
eigenstates which hints at the avoidance of the local
curvature singularity and 2. that the Quantum Einstein Equations are
non -- singular which hints at the avoidance of the global  
initial singularity. 
This rests on 1. a key technique developed for LQG and 2. the fact 
that there are no inhomogeneous excitations. We display the result of a 
calculation for LQG which proves that 
the (analogon of the) inverse scale factor, while densely defined, is 
{\it not} bounded from above on zero volume eigenstates. Thus, in full 
LQG, if curvature singularity avoidance is
realized, then not in this simple way.

In fact, it turns out that the boundedness of the inverse scale factor is 
neither necessary nor sufficient for curvature singularity avoidance and 
that non -- singular evolution equations are neither necessary nor 
sufficient for initial singularity avoidance because none of these 
criteria are formulated in terms of observable quantities. 
After outlining what would be required, we present the results of a 
calculation for LQG which could be a first indication that our 
criteria at least for curvature singularity avoidance are satisfied in 
LQG.
\end{abstract}

\section{Introduction}
\label{s1}

Loop Quantum Gravity (LQG) is a candidate for a Quantum Field Theory (QFT)
in four dimensions which achieves to unify the principles of Quantum 
Theory (QT) and General Relativity (GR). This means that LQG implements 
the fundamental feature of GR, its background independence, in a quantum
setting, see e.g. \cite{1} for books and \cite{1a} for reviews. 

To see how radical that ambition is, recall that ordinary QFT 
relies on a background spacetime $(M,g_0)$. Here $M$ is a differential
manifold and $g_0$ is a prescribed background metric thereon. Ordinary QFT 
now axiomatically assumes causality with respect to $g_0$: Given two 
smeared field operators 
$\phi(f),\;\phi(f')$ such that the supports of the smearing functions 
$f,f'$ are spacelike separated (the non -- spacelike geodesics with 
respect to $g_0$ starting from supp$(f)$ never hit supp$(f')$ and vice 
versa) then the field operators must (anti)commute. Thus we see that 
the background metric $g_0$ pivotally finds its way into the very 
definition of the field algebra. Without $g_0$ we therefore do not even
know what a quantum field is (see however \cite{2} for an interesting 
reformulation of ordinary QFT which may get us rid of $g_0$). 

The background $g_0$, however, is incompatible with GR, in particular
Einstein's equations, which state that the metric $g$ is a dynamical field 
which cannot be prescribed but must be determined dynamically according
to the matter energy momentum density. Hence, ordinary QFT  
neglects backreaction effects from matter on geometry (which is 
appropriate in scattering situations of negible particle numbers). 
Worse, if $g$ becomes itself a quantum field, in particular if there 
is no background $g_0$ such that the fluctuations $g-g_0$ are small  
in an appropriate sense, then ordinary QFT fails to be a valid description 
of the physical processes\footnote{The alternative, to quantize the 
fluctuations themselves and to construct a graviton QFT on $g_0$ fails 
due to 1. the non -- renormalizability of GR (see however \cite{3} 
for ideas concerning a non -- trivial UV fix point) and 2. 
in violent situations 
there may simply be no $g_0$ at all such that the fluctuations are 
small.}. It is precisely in those situations that a 
full fledged quantum theory of both matter and geometry must take over 
and it is almost granted that these situations arise for instance in the 
early universe. 

Thus, the task of a quantum gravity theory must be to define a quantum 
field just on a differential manifold $M$ and not on a background 
spacetime $(M,g_0)$. This means that one has to throw many aspects of 
conventional QFT over board. LQG is trying to precisely achieve this goal. 
As one might expect, given that already classical GR is a rather difficult
classical field theory, this is a very hard enterprise, both conceptually 
and technically. Despite these difficulties, LQG has made steady progress 
over the past fifteen years and 
we are getting closer and closer 
to being able to answer the question 
whether LQG really is a QFT of GR in the sense that 1. classical GR is the 
classical limit of LQG and 2. when the fluctuations of the quantum metric 
are small, then the framework of QFT on $(M,g_0)$ is recovered. In other
words, if these questions could be answered affirmatively, then we would
have a viable quantum gravity theory in front of us and we could start 
drawing conclusions and physical predictions from it and hopefully compare 
those with experiments.

However, even before we (hopefully) complete the construction of full LQG, 
it is possible and mandatory to test the theory by models which are 
conceptually and technically easier to handle and which still capture 
enough of the features of the full theory. To give an example, the 
hydrogen atom can be analyzed just using quantum mechanics for the 
electrons orbiting a classical nucleus and one gets fantastically close 
to the experiment. Hence, this gives us an important first idea. However, 
the full problem of the hydrogen atom must certainly
take into account the full QFT of the standard model\footnote{Even 
today that has not yet been done fully, usually one only considers QED 
effects but not e.g. the QCD physics of the proton.} and only then one 
can determine the Feinstruktur of the hydrogen atom energy spectrum such 
as the Zitterbewegung of the electron due to the interaction with the 
electromagnetic field. 

It is remarkable and important that the predictions
of the simple quantum mechanical computation are confirmed by the full 
theory because a priori it could be that the full QED calculation 
completely changes the picture. Of course, in the case of the hydrogen 
atom one has access to experimental data and so this was not expected if 
QED has something to do with nature. However, in case of absence of data
there is no reason why one should trust a toy model calculation at all.
In particular, it is unclear why a quantum mechanical calculation
describing only a finite number of degrees of freedom will reproduce the 
QFT aspects of the full problem involving an infinite number of degrees of 
freedom. One hopes of course that the full problem will only yield 
corrections to the model, however, that is far from granted and must be 
proved rigorously.

As far as LQG is concerned, in recent years a new class of toy models 
have been constructed which were coined Loop Quantum Cosmology (LQC).
These developments are mainly due to Bojowald, see e.g. the 
beautiful review 
\cite{4}. These models consider the truncation of the infinite dimensional 
phase space of GR to spatially homogeneous situations which is finite 
dimensional. Such truncations have been considered before and are called 
minisuperspace models \cite{5}. Traditionally they were quantized using 
the Schr\"odinger representation of the canonical commutation relations.
What is new in Bojowald's work is that he employs a different 
representation of the canonical commutation relations. This might come as 
a surprise because the Stone von Neumann theorem gives us a uniqueness 
result concerning the representation theory of the Weyl algebra for 
finite dimensional phase spaces. The catch
is that the Stone von Neumann theorem assumes, among other things, (weak) 
continuity of the Weyl operators. Bojowald drops this assumption because 
in full LQG the unique \cite{5b} representation \cite{5a} is also 
discontinuous, hence this 
representation is a much better model for LQG than the usual Schr\"odinger
representation. This kind of representations were previously discussed by 
e.g. Thirring \cite{6} and while exotic are used e.g. in solid state 
physics \cite{7}. Moreover, using those one can circumvent negative norm 
states in Maxwell theory \cite{8} or string theory which then 
does not lead to restrictions on the spacetime dimension \cite{9}.

A second important ingredient in Bojowald's work is a key technique 
discovered for full LQG in \cite{10} which allows to define certain 
operators densely on the LQG Hilbert space although their classical 
counterpart is ill defined when the metric becomes degenerate. Prime 
examples for such classically singular situations are cosmological big 
bang singularities. This key technique was used in \cite{10} in order to 
give, for the 
first time, a mathematical meaning to the Quantum Einstein Equations in 
full LQG, mathematically speaking, the so -- called Wheeler -- DeWitt 
equations could be defined densely on the LQG Hilbert space. Moreover,
these equations can be solved explicitly by following a constructive 
algorithm \cite{10a}. This works for arbitrary matter coupling \cite{10b}
and was tested successfully in 2+1 gravity \cite{10c}. 
For a modern version 
of this framework see \cite{11}. A simpler setting in which this key 
technique was used is the construction of the length operator in full LQG
\cite{11a}. This operator is also well (i.e. densely) defined, it is an 
unbounded but positive self -- adjoint operator whose spectrum is 
entirely discrete. 

Of course, due to the complexity of the full theory, LQG cannot be solved 
completely. It is precisely the virtue of toy models such as LQC that 
their mathematics is comparatively simple so that one can focus 
immediately on  
the conceptual analysis. Bojowald therefore could go very far in the 
quantization of the cosmological models although some important aspects 
such as the physical inner product are still missing. The most 
spectacular finding is that in LQC the operator corresponding to the 
inverse scale factor(s)\footnote{One for the isotropic models, three for 
the diagonal models.} is (are) bounded from above \cite{12,13a}. 
This hints at the following: Recall that in classical 
GR one characterizes singularities by global and local criteria 
\cite{17a}. The global criterion is the causal geodesic incompleteness 
of a (inextendible) spacetime. Equivalently,
the Einstein equations on 
globally hyperbolic spactimes with manifolds diffeomorphic to 
$\Rl\times \sigma$ cannot be solved for all values of $t\in \Rl$.
The local criterion is defined as 
the divergence of polynomial scalars built out of the Riemann tensor
and is used to characterize a global singularity in more detail. The local 
criterion by itself is unsatisfactory because a 
spacetime can have a global singularity while all curvature scalars
are regular. Notice that both criteria are to be measured by an
obeserver and in this sense are to be formulated in terms of observables 
(measurable quantities).\\ 
In classical cosmology both types of singularities are present.
Since in cosmology curvature scalars are 
given by powers of the inverse scale factor, its boundedness hints at 
the absence of the local singularity in LQC in the sense that
with respect to any kinematical state the curvature expectation value 
can never diverge. Moreover, the 
truncated Quantum Einstein Equations (Wheeler DeWitt Equation)
do not become singular at the classical Big Bang and 
one therefore can go before the Big Bang. This hints 
at global singularity avoidance in LQC.
Further consequences of the 
boundedness of the inverse scale factor are that the quantum geometry 
behaves very non -- classically at the classical singularity implying 
effective modifications of the Friedmann equations which then drive {\it 
geometric inflation without the necessity of fine tuning the inflaton
potential} \cite{13}. These results related to the boundedness of the 
inverse scale factor in LQC are so promising and spectacular that it was 
extensively highlighted in the
physics literature and public press. In particular, the boundedness
was mentioned as crucial to establish the non singularity of the quantum 
evolution equations, see e.g. \cite{Bojowald1}.

In \cite{14} it was shown that it is precisely those two ingredients
imported from full LQG into LQC,
1. the non -- standard representation of the canonical commutation 
relations and 2. the key technique \cite{10} to densely define classically 
degenerate operators, which are responsible for the absence of the 
local singularity in LQC and similar models inspired more by a 
metric than 
connection based approach. However, as outlined above, on general grounds
one must critically examine the stability of such model calculations when
one reintroduces the field theory degrees of freedom. This has been 
stressed before for geometrodynamics in \cite{14a} where one generically 
finds that the model even quantitatively has large deviations from the 
full theory or more complete theory. More recently also LQC has been 
examined critically   
\cite{14b} concerning the issue of the semiclassical limit but here the 
results seem to be inconclusive so far. 

The main purpose of this paper 
is to contribute to the stability analysis of LQC versus LQG.\\
\\
The paper is organized as follows:\\
\\ 
\\
In section two we summarize and compare the main features of LQG and LQC
for the unfamiliar reader. Experts can safely skip this section.\\
\\
In section three we report the results of a computation carried out in our 
companion paper \cite{16} within full LQG which 
is analogous to the one carried out in LQC. In LQC this calculation proves 
the boundedness of the inverse scale factor. 
More precisely, we ask, as in LQC, whether the
part of the energy momentum operator which probes the quantum metric is 
bounded from above on the kinematical Hilbert space of full LQG, at 
least at the classical singularity, that is, when the volume of space 
vanishes. This precisely mirrors the calculation of \cite{12}.
The result is negative: {\bf In full LQG the (analogon of) the inverse 
scale factor is unbounded from above even when the quantum volume 
vanishes.} 
This is true even when we compute its norm with respect to a 
state of zero volume which is homogeneous and isotropic on large 
scales. This proves that the boundedness of the inverse scale factor
in isotropic and homogeneous LQC does not extend to the full theory even
when restricting LQG to those states which one would use to
describe a maximally homogeneous and isotropic situation (modulo
fluctuations). 

It is true that in non -- isotropic, homogeneous
models the inverse scale factor is also unbounded on the full Hilbert
space. However, in those more general models it is still bounded on zero
volume eigenstates which again seems to indicate that the local 
singularity is evaded in LQC. Our result is stronger: No matter whether 
homogeneous
and/or isotropic, spherically symmetric, cylindrically symmetric etc., the
inverse scale factor is unbounded on zero volume
eigenstates in full LQG and on the full Hilbert space anyway. The reason 
for this
is that LQG, in contrast to LQC, admits inhomogeneous, microscopical
excitations which precisely account for the
unboundedness. We will see this explicitly when we construct the
zero volume eigenstates on which the inverse scale factor has
arbitrarily large norm. Notice that
meanwhile also inhomogeneous models such as spherically symmetric
or cylindrically symmetric ones have been quantized by LQG methods in which
the inverse scale factor apparently also stays bounded 
\cite{Bojowald}.
While these do have local, inhomogeneous degrees of freedom
(in the spherically symmetric case without matter, at least before
solving the constraints),
as our calculation reveals, the inhomogeneous excitations of
full LQG are more general than in those models and thus give unboundedness
even on states of LQG describing a sector appropriate for those 
more general models. \\
\\
However, this does not mean that LQG does not predict the absence of the
local initial singularity. Namely, in section four we remark that 
an operator may be unbounded but still it may be bounded when restricted 
to a subspace of the Hilbert space. Thus, the boundedness of the inverse 
scale factor is not a necessary criterion for local singularity avoidance. 
In our case we are interested in 
a sector which describes a collapsing universe which is homogeneous and 
isotropic at large scales. We perform a corresponding calculation
which may be taken as an indication that 
the expectation value of the analogon of the inverse scale factor in LQG, 
with respect to kinematical\footnote{Kinematical states, in contrast to 
physical states, do not solve the Quantum Einstein Equations. 
Non -- gauge invariant operators such as the inverse scale factor can 
only be probed on the kinematical Hilbert space.}, 
coherent states \cite{20} peaked on homogeneous and 
isotropic initail data, {\bf is bounded from above at the 
Big Bang}\footnote{Notice that coherent states peaked
on homogeneous and isotropic initial data still have inhomogeneous and 
anisotropic excitations, however, they are small.}. We did this for scalar 
matter only but qualitatively nothing changes in more general 
situations. Also our result is completely general: We derive a bound of 
the inverse scale factor on coherent states peaked on an arbitrary point 
in phase space. This formula can then be specialized also to inhomogeneous 
models and for the Bianchi I (Kasner) case it is easy to see that one gets 
boundedness as well.\\
\\
In section five we stress that the results of section four are promising 
but 
inconclusive: Unfortunately, the boundedness of the inverse 
scale factor is also not a sufficient criterion for 
local singularity avoidance. 
This has to do with the fact that both singularity criteria 
fundamentally have to be discussed at the level of physical 
observables and physical states. They must be be disussed 
separately because the quantum theory could avoid one of them but not the
other.\\ 
Concerning the local singularity, the inverse scale factor is not gauge 
invariant and therefore not an observable. It can be turned into an 
observable using the technique of partial observables \cite{15,15a} but 
then its spectrum on the physical Hilbert space can differ drastically 
from the kinematical spectrum \cite{20a}. Hence, reliable
statements about the local singularity can only be obtained on the 
physical Hilbert space.\\ 
As far as the global singularity is concerned, physical states 
automatically solve the quantum Einstein equations so there seems to be 
no sign of the global singularity at the level of physical states, by 
construction\footnote{Of course, the inverse scale factor must be densely 
defined to even define the Quantum Einstein Equations but that has been 
established for LQG in \cite{10,10b}.}. More precisely, in order to solve 
the Quantum Einstein Equations, one takes an arbitrary superposition 
of kinematical states and then obtains a recursion relation for the 
coefficients. These have been derived explicitly for full LQG in 
\cite{10a} and those of LQC \cite{12} are similar but simpler.
One of the labels of these coefficients can be identified with an 
unphysical time parameter. If one chooses that label in such a way 
that its range is 
(a discrete subset of) the full real axis, a solution to the Quantum 
Einstein Equations, that is, a physical state,
describes an entire quantum universe (a history) and involves the full 
range of the unphysical time parameter. {\it Thus there is no sign of the 
global singularity within a given solution even in the full theory}.
However, following \cite{12} one may argue that the existence
of a global singularity
could manifest itself in the fact that the recursion relations break down
at zero volume. Such a breakdown can manifest itself in the following 
two ways:\\
A.\\ 
It can lead
to a restriction or additional consistency condition on the
freedom in the choice of the initial
data (coefficients) that parameterize the solution space.
This makes the solution space smaller than it would be otherwise.
In particular, if the
number of additional conditions is larger than the free initial parameters
of the solution, the solution space would be empty.\\
B.\\
It can happen that certain coefficients remain undetermined by the
recursion relation and lead to an indeterministic quantum evolution
(given initial coefficients, we cannot construct the full solution
without making additional choices). This makes the solution space larger
than it would be otherwise. Notice that this undeterministic quantum
evolution corresponds to an undeterministic classical evolution (initial
value problem). However, by itself such ambiguities are not physically
relevant because this is not an evolution of physical quantities. Even 
classically it is an evolution with respect to an unphysical time of 
unphysical quantities. In terms of physical (gauge invariant) quantities 
{\it there is no unphysical evolution at all by definition} because Dirac
observables Poisson commute with the Hamiltonian constraint. Hence, 
what we actually have to do in order to observe a possible breakdown of 
evolution is to
consider an evolution with respect to a physical Hamiltonian
and of physical observables.
We will discuss this at length in section five.\\
We now point
out that absence of either effect A or B is neither a sufficient nor a 
necessary criterion for global singularity avoidance.\\
To see that it is not necessary, notice that the presence of either effect
indeed affects the size of the set of formal solutions to
the evolution equations. However, even if such effects
exist, it may still be the case that the physical Hilbert space is large
enough in order to accomodate all semiclassical physical states which
describe all
classical spacetimes (at large scales). To see that it is not
sufficient, notice that not
all formal solutions to the Quantum Einstein Equations correspond to
physical states because they might not be normalizable with respect to
the physical inner product or they might have zero norm.
In both cases we must drop them from the set of physical states.
Hence, even if effects A or B are absent, it could turn out, in the
worst case, that most of the solutions that one finds do not lie
in the physical Hilbert space and then again one would be confronted with
a global singularity because there could be not even be a semiclassical 
sector
describing cosmology. Furthermore, as argued above, global singularity 
avoidance 
must be formulated in terms of physical observables which has 
not been done yet. Therefore, the breakdown 
of the recursion or the absence thereof is inconclusive for the presence 
or absence of global singularities 
unless one knows the physical Hilbert space and the corresponding Dirac 
observables.
 
We conclude: {\bf In order to decide on the presence or absence of
both the local and global singularity, detailed knowledge of the physical 
Hibert space is a necessary 
prerequisite.}. This Hilbert space is known to 
exist both in LQG and LQC, following for instance \cite{11}. However,
the corresponding physical inner product is rather difficult to 
construct explicitly,
for any of the current versions of the Hamiltonian constraint,
which is why a definitive conclusion on the singularity issue
is not available at the moment.

To improve on this, we first of all broadly outline which steps should be 
fundamentally performed 
in order to prove the absence of both the global and  
local Big Bang singularity in full LQG. We give both a precise and 
an approximate scheme. 
The ideal (precise) procedure involves the exact
knowledge of the Hilbert space of physical states and a sufficiently 
large set of Dirac observables (gauge invariant quantities) which, as we 
said, are not 
yet available neither in full LQG nor in LQC in 
a form explicit enough. The
approximate scheme stays within the kinematical Hilbert space over which 
one has excellent control. The calculation performed in section 
four can be viewed as a simplified version of the approximate scheme. 
The application of either scheme is left for future research and again 
LQC would be an ideal testing ground for this scheme.\\
\\
We conclude in section six summarizing the status of the absence of 
the initial singularity in full LQG. Notice that qualitatively all we have 
said is also applicable to other singularities such as those connected 
with black holes \cite{20d,20c}. In fact, most considerations within LQG 
concerning black holes are based on 
minisuperspace calculations which exploit the fact that the interior of a 
black hole can be mapped into a homogeneous Kantowski -- Sachs model
so that one can transfer, almost literally, the LQC results. However, our 
remarks prevail: No reliable conclusions can be drawn before one works 
with physical operators and physical states of the full theory. This has 
also been stressed in \cite{20c}.\\
\\
Notice that this paper is not meant as a criticism of LQC by 
itself, in the contrary, LQC has provided us with many fruitful new 
conceptual ideas and one would hope that its results hold also LQG, albeit
in a technically different realization.
However, we want to draw attention to the fact that LQC is far from being 
full LQG and that therefore caution should be applied when trying to 
extrapolate LQC results to LQG. This extends also to more general reduced
models such as the spherically symmetric or cylindrically symmetric ones.

\section{Relation between LQG and LQC}
\label{s0}

For the benefit of the reader unfamiliar with the basic structure of 
either LQG or LQC we here summarize elements of both frameworks and 
compare them. The presentation will be oversimplified and the 
mathematical details will be discared.
As we will see, LQC does not describe a stable subsector
of LQG, rather it is a truncation in which all but finitely many degrees 
of 
freedom are cancelled by hand. Experts can safely skip this section.

\subsection{Elements of LQG}
\label{s0.1}

The classical starting point of LQG is a Hamiltonian formulation of the 
Einstein Hilbert action, however, written in unusual variables. Instead
of using the three metric $q_{ab}$ on the spatial slices $\sigma$ of a 
foliation of the four manifold $M\cong \Rl\times \sigma$ and the extrinsic 
curvature $K_{ab}$ as the canonical variables as advocated by Arnowitt,
Deser and Misner, one uses variables that are more famililar from the 
canonical formulation of Yang Mills theories. They consist of an 
$SU(2)$ connection $A_a^j$ and an electric field $E^a_j$. Here 
$a,b,c,..=1,2,3$ are tenorial indices and $j,k,l,..=1,2,3$ are $su(2)$
indices. The origin of $SU(2)$ is that $SU(2)$ is the universal covering 
group of $SO(3)$ (necessary for coupling of spinorial matter) and $SO(3)$
appears naturally when we write the three metric in terms of cotriads
(or frame fields) 
$e_a^j$, that is, $q_{ab}=e_a^j e_b^k \delta_{jk}$. Indeed, the cotriads 
are determined by $q_{ab}$ only up to an $SO(3)$ rotation and this is 
how $SU(2)$ finds its way into this formulation as a gauge group. 

The relation between $(A,E)$ and $(q,K)$ is 
\be \label{0.1}
A_a^j=\Gamma_a^j+\beta K_{ab} e^b_j,\;\;E^a_j=\sqrt{\det(q)} e^a_j
\ee
Here $\Gamma$ is the spin connection (a certain function of $e_a^j$ and 
its first spatial derivatives), the triad $e^a_j$ is the inverse of 
$e_a^j$, i.e. 
$e^a_j e^j_b=\delta^a_b,\;\; e^a_j e^k_a=\delta^k_j$ and $\beta>0$ is a 
parameter called the Immirzi parameter. That $(A,E)$ form a canonical pair 
means that
\be \label{0.2}
\{E^a_j(x),A_b^k(y)\}=8\pi G \beta \;\delta^a_b\;\delta^k_j\;\delta(x,y)
\ee
all others vanishing. The phase space coordinatized by $(A,E)$ is subject 
to the usual spatial diffeomorphism constraint which in these variables 
can be written as
\be \label{0.3}
C_a=F_{ab}^j\; E^b_j \mbox{ where } F_{ab}^j=\partial_a A_b^j-\partial_b 
A_a^j+\epsilon_{jkl} A_a^k A_b^l
\ee
is the curvature of $A$, the Hamiltonian constraint
\be \label{0.4}
C=\frac{\epsilon_{jkl} [F_{ab}^j-(1+\beta^2) \epsilon_{jmn} K_a^m K_b^n]
E^a_k E^b_l}{\sqrt{|\det(E)|}}
\ee
where $K_a^j=A_a^j-\Gamma_a^j(E)$ and the additional Gauss constraint
\be \label{0.5}
C_j=\partial_a E^a_j+\epsilon_{jkl} A_a^k E^a_l
\ee
which gets us rid of the $SU(2)$ degrees of freedom. We see that GR can be 
cast into the form of a $SU(2)$ Yang Mills theory (which is also subject 
to $C_j=0$) with one important difference: The background dependent Yang 
Mills Hamiltonian density
\be \label{0.6}
H=\frac{q^0_{ab}}{\sqrt{\det(q^0)}}\;[E^a_j E^b_k+B^a_j B^b_k]\delta^{jk}
\ee
where $B^a_j=\epsilon^{abc} F^j_{bc}/2$ is the magnetic field and
$q^0$ is the spatial projection of the background metric $g^0$,
is replaced by the background independent constraints $C_a,C$. 

If there is matter present such as a scalar field $\phi$ then the 
constraints $C_a,\; C$ are augmented by additional terms
\be \label{0.7}
C^{scalar}_a=\pi \phi_{,a} \mbox{ and } 
C^{scalar}=\frac{1}{2}\frac{\pi^2+E^a_j E^b_j 
\phi_{,a} \phi_{,b}}{\sqrt{|\det(E)|}}+\sqrt{|\det(E)|} V(\phi)
\ee
where $\pi$ is the momentum conjugate to $\phi$. 
 
Canonical quantization now, very roughly, means to find a Hilbert space 
${\cal H}_{Kin}$ on which the canonical Poisson brackets (\ref{0.2})
are implemented as canonical commutation relations and to impose the 
quantum constraints, e.g. $\hat{C}\psi=0$. The resulting space of 
solutions is the physical Hilbert space on which the (Dirac) observables,
that is, the gauge invariant functions $F$ on phase space satisfying
$\{C_a(x),F\}=\{C(x),F\}=\{C_j(x),F\}=0$ act as self adjoint operators
$\hat{F}$. This simple recipe has to be refined in QFT because of the 
singular short distance behaviour of the fields $A,E$. In order to deal 
with this problem one has to smear the fields. There are many ways to smear 
the fields, each of which gives rise to a different algebra 
$\mathfrak{A}$ of elementary
observables\footnote{They are called elementary because one can write 
every function on phase space in terms of (limits of) them. They are 
not physically observable because they are not gauge invariant.}. Guided 
by experience with canonical lattice gauge field theory in LQG one 
considers the following objects:\\ 
1. For each curve $e$ in $\sigma$
the holonomy of the connection $A$ along $e$, that is,
\be \label{0.8}
A(e):={\cal P}\;\exp(\int_e A):=1_2+\sum_{n=1}^\infty
\int_0^1\; dt_1\; \int_{t_1}^1\;dt_2\; ..\;\int_{t_{n-1}}^1\; dt_n\;
A(t_1)..A(t_n) 
\ee
where $e:\;[0,1] \to e; t \mapsto e(t)$ is a parameterization of the 
curve and $A(t):=A_a^j(e(t)) \dot{e}^a(t) \tau_j$ with 
$\tau_j=-i\sigma_j/2$ a basis of $su(2)$ ($\sigma_j$ are the Pauli 
matrices).\\
2. For each surface $S$ and each $su(2)$ valued function $x\mapsto f^j(x)$
the electric flux of $E$ through $S$
\be \label{0.9}
E_f(S)=\int_s\; d^2u\; f^j(X_S(u))\; \epsilon_{abc} E^a_j(X_S(u))\;
X^b_{S,u^1}(u)\; X^c_{S,u^2}(u)
\ee
where $X_S:\;s\subset \Rl^2\to S;\;u=(u^1,u^2)\mapsto X_S(u)$ is a 
parameterization of 
$S$. \\
Using (\ref{0.2}) one can show that the Poisson brackets\footnote{More 
precisely the corresponding Hamiltonian vector fields.} of the holonomies 
and fluxes close among each other so that they define a well defined 
algebra $\mathfrak{A}$.  

The kinematical Hilbert space of LQG can now be described very easily as 
follows:\\
A union of curves $e$ forms a graph $\gamma$. By subdividing the curves we 
may assume that the curves are mutually disjoint except for the endpoints.
These curves are then called edges of $\gamma$ which intersect in the 
vertices. We will denote by $E(\gamma),\;V(\gamma)$ the set of 
edges and vertices of a graph respectively. The wave functions are then 
of the form 
\be \label{0.10}
\psi(A)=\psi_\gamma(A(e_1),..,A(e_N))
\ee
where $\gamma$ can be any graph, $N$ is the number of edges of the graph 
$\gamma$ and $f_\gamma:\;SU(2)^N \to \Cl$ is a complex valued function 
on $N$ copies of $SU(2)$. \\
The holonomy operator acts by multiplication
\be \label{0.11}
[\widehat{A(e)}\; \psi](A):=A(e)\; \psi(A)
\ee
while the fluxes act by derivation
\be \label{0.12}
[\widehat{E_f(S)}\; \psi](A):=i\hbar \{E_f(S),\psi(A)\}
\ee
The scalar product between functions $\psi,\psi'$ defined via (\ref{0.10})
over graphs $\gamma,\gamma'$ respectively is defined as follows:
Take the union graph $\gamma^\dprime:=\gamma\cup\gamma'$ and write 
the edges $e,e'$ of $\gamma,\gamma'$ respectively as compositions 
of edges $e^\dprime$ of $\gamma^\dprime$, say $e=e^\dprime_1\circ ..\circ
e^\dprime_n$. Now use the fact that the holonomy factorizes, e.g. 
$A(e)=A(e^\dprime_1)..A(e^\dprime_n)$ to write $\psi,\psi'$ as functions 
over $\gamma^\dprime$. Then
\be \label{0.13}
<\psi,\psi'>:=\int_{SU(2)^N}\;d\mu_H(h_1)\;..\;d\mu_H(h_N)\;
\overline{\psi_{\gamma^\dprime}(h_1,..,h_N)} \;
\psi'_{\gamma^\dprime}(h_1,..,h_N) 
\ee
where $N$ is the number of edges of $\gamma^\dprime$ and $\mu_H$ is the 
Haar measure on $SU(2)$. Specifically, if we write 
$h=\cos(\chi) 1_2+\tau_j n^j(\theta,\varphi) \sin(\chi)$ with 
$\chi,\theta\in [0,\pi)$ and $\varphi\in [0,2\pi)$ where \\
$\vec{n}=(\sin(\theta) 
\cos(\varphi),\sin(\theta)\sin(\varphi),\cos(\theta))$ is the usual unit 
vector on the sphere then $d\mu_H(h)=c\sin^2(\chi)\sin(\theta) 
d\chi\;d\theta\;d\varphi$ and $c$ is a normalization constant fixed 
by requiring that $\mu_H(SU(2))=1$.

It turns out that the Hilbert space ${\cal H}_{Kin}$ has a convenient
orthonormal basis consisting of the spin network functions (SNF) 
$T_{\gamma,\vec{j},\vec{I}}$. For the purpose of this paper 
it is enough to say that they are specific functions of the type 
(\ref{0.10}) which are polynomials in the holonomies $A(e),\;e\in 
E(\gamma)$. They are labelled by a graph $\gamma$, a collection 
$\vec{j}:=\{j_e\}_{e\in E(\gamma)}$ of spin quantum numbers 
$j_e,\;2j_e=1,2,3,..$, one for each edge, and a collection of other 
quantum numbers $\vec{I}=\{I_v\}_{v\in V(\gamma)}$ called intertwiners, 
one for each vertex. Since the label $\gamma$ is continuous, the basis is 
uncountable and the Hilbert space is therefore not separable. We will 
use the short hand $s=(\gamma(s),\vec{j}(s),\vec{I}(s))$ for a spin 
network label.

Let us now turn to the quantum constraints. The Gauss constraint imposes 
certain restrictions on the quantum numbers $\vec{I}$ of the SNF and is 
easily taken care of. The spatial diffeomorphism constraint enforces that
we should identify all spin network labels $s,s'$ if there exists a 
spatial diffeomorphism $\varphi$ of $\sigma$ such that
$\varphi(\gamma(s))=\gamma(s')$ and so is also easily taken care of.
The real challenge for LQG (and any other quantum gravity theory) is the 
Hamiltonian constraint. Looking at (\ref{0.4}) the mere definition of
the operator $\hat{C}$ looks like a hopeless task due to the non -- 
linearity of the constraint. It is here where the key technique of 
\cite{10} comes into play:\\ 
It is easy to see that the nonlinear expression involving the electric 
field $E$ can be written in the compact Poisson bracket form
\be \label{0.14}
e_a^j=\frac{\epsilon_{abc} \epsilon^{jkl} E^b_k E^c_l}{\sqrt{|\det(E)|}}
\propto \{A_a^j,V\} \mbox{ where } V=\int_\sigma d^3x \sqrt{|\det(E)|}
\ee
is the volume\footnote{In the case of $k=0$ this diverges classically.
However, we only need its Poisson bracket which is well defined.} of 
$\sigma$. The advantage of this way of writing 
(\ref{0.14}) is that there exists a volume operarator corresponding to $V$
which can be diagonalized in terms of spin network states. Thus, 
(\ref{0.14}) can be quantized by substituting Poisson brackets by 
commutators divided by $i\hbar$. Next, consider the smeared quantity
$C(N)=\int d^3x N(x) C(x)$ where $N$ is a test function. We may write this 
as $C(N)=C_F(N)+C_K(N)$ where $C_F,C_K$ respectively correspond to the 
terms in (\ref{0.4}) involving the term $F_{ab}^j$ or $K_a^m K_b^n$ 
respectively. Now it is easy to see that 
\be \label{0.15}
K_a^j(x)\propto \{A_a^j(x),\{C_F(1),V\}\}
\ee
where $C_F(1):=C_F(N)_{N=1}$. It follows that we know how to quantize 
$C(N)$ once we know how to quantize $C_F(N)$\footnote{The condition 
$C(x)=0$ for all $x$ is equivalent to requiring $C(N)=0$ for all $N$ 
because the support of $N$ can be localized arbitrarily.}. In order to do 
this it remains to write $C_F(N)$ in terms of holonomies. This is achieved 
by recalling that for a curve $e^a_{x,u}(t):=x^a+t u^a$ and loop 
$\alpha_{x,u,v}(\epsilon):=
e_{x,u}(\epsilon) \circ
e_{x+\epsilon u,v}(\epsilon) \circ
e_{x+\epsilon(u+v),u}(\epsilon)^{-1} \circ
e_{x+\epsilon v,v}(\epsilon)^{-1}$
respectively we have the expansions 
$A(e_{x,u}(\epsilon))=1_2+\epsilon u^a A^j_a(x) \tau_j/2+O(\epsilon^2)$ 
and $A(\alpha_{x,u,v}(\epsilon))=1_2+\epsilon^2 u^a v^b F_{ab}^j(x) 
\tau_j/2+O(\epsilon^3)$. 
Therefore, if we write the integral $C_F(N)$ as the limit of a Riemann sum 
over boxes of coordinate volume $\epsilon^3$ then in the limit it is 
appropriate to replace $F_{ab}^j$ by a holonomy along a loop on the 
boundary of the box while the $A_a^j$ in the Poisson bracket $\{A_a^j,V\}$
may be replaced by a holonomy alon an edge on the boundary of that box.

The details of taking that limit can be inferred from \cite{10,11}.
Roughly speaking, one defines (a spatially diffeomorphism invariant 
version of) the constraint on the space of solutions to the diffeomorphism
constraint on which the limit becomes trivial because what matters to the 
constraint are only the terms of the Riemann sum next neighbouring the 
vertices and those terms are diffeomorphic to each other when we refine 
the Riemann sum. 

\subsection{Elements of LQC and Comparison with LQG}
\label{s0.2}

LQC, by definition, is the canonical quantization of the cosmological 
truncation of classical GR by the methods of LQG\footnote{Originally 
one tried to define LQG as a symmetric sector of distributions on a dense 
sunspace of the LQG Hilbert space together with the dual action of the 
full LQG operators. However, this dual action does not preserve the 
symmetric sector. See also the end of this section. We adopt here the 
correct and modern point of view spelled out in \cite{17ba}.}. 
Hence one first 
specializes the functions $A_a^j(x), E^a_j(x)$ to the spatially 
homogeneous case, plugs those into the formulae for the holonomies and
fluxes and then computes the corresponding algebra $\mathfrak{A}_{cosmo}$. 
That algebra is of course tremendously smaller than the full algebra 
$\mathfrak{A}$ but one quatizes it in analogy to LQG, that is, one chooses 
a Hilbert space representation which is as similar as 
possible to the one of LQG. Finally, one plugs the simplified expression 
for $(A,E)$ into the constraints and quantizes them as in LQG, in 
particular one uses the technique (\ref{0.14}) which enables to 
avoid denominators (which are singular at the classical singularity)
at the price of employing Poisson brackets with the volume functional.\\
\\
Let us see this in more detail\footnote{What follows is a drastically 
simplified version of \cite{17ba}.}:\\
We will consider the simplest case: The isotropic, flat FRW cosmologies
for which the line element reads
\be \label{0.16a}
ds^2=-dt^2+a(t)^2 d\vec{x}^2
\ee
where $a$ is the scale factor. Here we take $t,a$ to have dimensions
of length.
One fixes both the spatial diffeomorphism constraint and the Gauss 
constraint by setting
\be \label{0.16}
A_a^j:=q \delta_a^j/3,\;\;E^a_j:=p\delta^a_j
\ee
where $p,q$ are spatial constants. It follows that $a=\sqrt{|p|}$ and 
that $\{p,q\}=8\pi \beta G$ are canonically conjugate. It is easy to see 
that with 
(\ref{0.16}) both the Gauss and spatial diffeomorphism constraint vanish
identically as they should and that the gravitational 
contribution to the Hamilonian constraint becomes 
$C=- \alpha q^2 \sqrt{|p|}$ with a positive constant $\alpha$.
For a (homogenous) 
scalar field we find
\be \label{0.17}
C^{scalar}\propto \frac{\pi^2}{2\sqrt{|p|}^3}+\sqrt{|p|}^3 V(\phi)
\ee
As one can show (see the end of section \ref{s4}) the classical 
singularity is encoded in the kinetic term only.

The holonomy and flux become respectively 
\be \label{0.18}
A(e)=\cos(q r_e)+\tau_j n^j_e \sin(r_e q),\;\;       
E_f(S)=[\int_S ds_a f^a] p
\ee
where $e^a(1)-e^a(0)=:r_e n^j_e,\;\;(n_e^j )^2=1$.
By varying $e$ and taking traces we may generate the functions 
$T_r:=\exp(i r q)$ where $r$ can be any real number. The reduced 
holonomy flux algebra therefore is generated by the $T_r$ and $p$.

The reduced Hilbert space is now obtained by asking that the $T_r$ form an 
orthonormal basis, i.e. $<T_r,T_{r'}>=\delta_{r,r'}$ so that they play the 
analogon of the SNF. Here the continuous graph label and the discrete spin 
label of the SNF have joined into one continuous label $r$.
The operators $\widehat{T_r},\;\hat{p}$ act by multiplication and 
derivation respectively, explicitly $\widehat{T_r}\;T_{r'}=T_{r+r'}$ and 
$\hat{p}\;T_r:=i\hbar\{p,T_r\}=-r\ell_p^2\beta  T_r$ where $\ell_p^2=
8\pi G \hbar$ is the Planck length squared.

Finally, in order to quantize objects like the inverse scale 
factor $1/a=1/\sqrt{|p|}$ we notice the 
classical identity 
\be \label{0.19} 
\frac{1}{\sqrt{|p|}}\propto \frac{T_{r_0} \{T_{-r_0},\sqrt{|p|}\} -T_{-r_0} 
\{T_{r_0},\sqrt{|p|}\}}{i 8 \pi \beta G r_0} 
\ee 
which holds for every 
$r_0\not=0$. Notice that $\sqrt{|p|}\propto V^{1/3}$ where $V$ is the 
volume\footnote{For $k=0$ this volume is actually infinite. We mean here 
the volume divided by the comoving volume.} so that (\ref{0.19})
is indeed analogous to (\ref{0.14}). Notice that the spectrum of the 
volume 
operator is essentially given by $|r|^{3/2}$ with $T_r$ as normalizable 
eigenfunctions (it takes a continuous range but consists only of 
eigenvalues, there is no continuous part). Hence the only zero volume 
state is $T_0=1$. It follows immediately that if the inverse scale factor 
can be densly defined on the $T_s$ then at zero volume it is bounded.  

Expression (\ref{0.19}) is readily quantized by substituting Poisson 
brackets by commutators divided by $i\hbar$. This gives
\be \label{0.20}
\widehat{\frac{1}{\sqrt{|p|}}} T_r 
\propto 
\frac{T_{r_0} \widehat{\sqrt{|p|}} T_{-r_0} 
-T_{-r_0} \widehat{\sqrt{|p|}} T_{r_0}}{\beta \ell_P^2 r_0}\;\;T_r
=\frac{\sqrt{|r+r_0|}-\sqrt{|r-r_0|}}{r_0 \sqrt{\beta} \ell_P} \;T_r
\ee
which is bounded (in fact vanishes) at $r=0$. It is even bounded for all 
$r$ and coincides with the operator $1/\sqrt{|\hat{p}|}$ for $r\gg r_0$.
The parameter $r_0$ is a quantization ambiguity, called a factor ordering 
ambiguity which is also present in LQG \cite{17c}
although it is there labelled by a 
discrete rather than continuous parameter.  Notice that by the constraint 
$C=C_{geo}+C_{scalar}=0$ we obtain $q^2\propto p^2/(2a^4)+V a^2$ which 
means that $q\propto a^{-2}$ at $a\to 0$ (see the discussion at the end of 
section \ref{s4}) so that $C_{geo}\propto a^{-3}$ classically. However,
quantum mechanically $C_{geo}$ is bounded at $a\to 0$ because $q$ is 
quantized by the bounded expression $(T_{r_0}-T_{-r_0})/(2i r_0)$. 

So far we have only discussed the isotropic model. In the non -- 
isotropic but diagonal model \cite{13a} similar remarks apply because 
the three scale factors mutually commute. Hence the inverse scale 
factors can be treated separately just as in the isotropic case.\\ 
\\
Let us summarize:\\
LQC is the usual cosmological minisuperspace phase space quantized by 
LQG methods. It borrows in an essential way the Hilbert space \cite{5a}
and the key technique \cite{10}. LQC therefore seems to to confirm 
the aspects \cite{5a,10} for LQG. 

However, despite these similarities there are 
important differences:\\
To begin with, LQC does not have a graph label. That is of 
course 
expected because in cosmology all points are equivalent due to 
homogeneity, hence there is only one point and thus no room for a 
graph.
Consequently, there is no such thing as the valence of a vertex.
These facts lead to an enormous simplification of the volume spectrum 
which is only modestly degenerate (multiplicity two for all
$r\not=0$, multiplicity one for $r=0$) while in LQG the volume spectrum is 
infinitely degenerate, in particular for zero eigenvalues (just apply a 
diffeomorphism to a given volume eigenstate which changes the graph). 
Moreover, the 
volume spectrum in LQG has discrete range while in LQC it has continuous 
range. Next, in LQG the Hamiltonian constraint modifies the graph of a 
spin network state while in LQC that is not possible since there is no 
graph. That leads to an enormous simplification concerning the 
computation of the kernel of the Hamiltonian constraint which boils down 
to the solution of a recursion relation within a finite 
dimensional parameter space while in LQG the corresponding parameter 
space \cite{10b} is infinite dimensional. 
Finally, while it is possible to interpret the LQC states as 
distributional LQG states \cite{17d} (simply multiply an ordinary LQG 
state by a 
$\delta$ distribution supported on homogeneous connections) this 
distributional space of states is not preserved by the (dual) 
action of the operators of 
LQG, hence LQC is not an invariant (distributional) sector of LQG.
It really is a model whose connection to LQG involves the classical 
truncation, thus inhomogeneous fluctuations are switched off by hand
rather than being quantum mechanically suppressed.\\
\\
We conclude: By definition of a toy model with a finite dimensional phase 
space,
LQC cannot model the QFT aspects of LQG. LQC models very successfully two 
aspects of LQG, namely those of \cite{5a,10}. However, it is far from 
granted that those predictions of LQC which are sensitive to the QFT  
aspects will hold in LQG as well. As we just saw, the spectrum of the 
inverse scale 
factor {\it is} sensitive to those aspects since the volume operator is. 
It is precisely one of the 
purposes of the present paper to investigate whether these differences 
lead to a qualitatively different picture in LQG when applied to 
cosmology.

\section{Unboundedness of the Analogon of the Inverse Scale Factor in 
Full LQG}
\label{s2}

We can explain the reason for the unboundedness of the operator 
concerning the inverse scale factor in very intuitive terms 
without going through the (admittedly lengthy) analysis. We will 
compare this to the isotropic LQC model, similar arguments apply to the 
the diagonal models:\\ 
The spectrum of the inverse scale factor can, roughly speaking, be 
obtained by taking the difference between neighbouring eigenvalues of the 
volume operator (discrete derivative). This is the case because, according 
to (\ref{0.14}) the 
inverse scale factor can be obtained from the Ko -- Dreibein as
$e=\{A,V\}$ where $V$ is the volume and $A$ is the connection. 
This is the 
key technique of \cite{10} as mentioned above. Upon quantization the 
connection gets replaced by a holonomy, that is, $\hat{e}\propto
h[h^{-1},\hat{V}]=V-h\hat{V} h^{-1}$ where $h$ is a the holonomy of $A$ 
along an edge. The holonomy changes the spin labels of a spin network 
state by one unit, hence we get a discrete derivative in the spin 
parameter space. Now the spectrum of the volume operator 
depends parametrically, among 
other things, precisely on the spins of the edges of a spin network 
attached to a vertex and additional intertwiner quantum numbers and this 
is the discrete derivative of the volume spectrum that we mentioned.

If we view the spectrum 
of the volume operator as a function of 
these parameters then in isotropic LQC this parameter space is 
one dimensional, it is a 
function $j\mapsto V(j)$ (in LQC the parameter $j:=r$ of the previous 
section takes continuous and not discrete values but that does not change 
the qualitative picture). In LQG on the 
other hand it is arbitrarily high dimensional, the dimensionality being 
governed by the number $n$ of vertices and the valences $N_v$ of the 
individual vertices, it is a function of the form
$(j^{(1)}_1,..,j^{(n)}_{N_n})\mapsto \sum_{v=1}^n 
V_v(j^{(v)}_1,..,j^{(v)}_{N_v})$ where 
the $n,N_v$ can be arbitrarily large. In 
both cases the functions $V$ are 
nowhere singular except when at least one of the $j$ tends to infinity.

The point is now that in isotropic LQC the zero volume condition
$V(j)=0$ has a finite number of solutions $j=j_0<\infty$ (in fact one).
The spectrum of 
the inverse scale factor at zero volume is then, roughly speaking, given 
by $V(j_0+\delta j)-V(j_0)$ (using suitable discrete units for the 
parameters $j$, in LQC $\delta j=r_0$) which is 
finite since $j_0<\infty$ whence we get boundedness 
at zero volume. In LQG on the other hand the situation is more 
interesting: The zero volume condition now gives an infinite number
(parameter space of codimension one) of solutions. Pictorially 
speaking, the spectrum 
of the volume operator plotted against the parameter space can be 
envisioned as a complicated landscape of mountains which is disrupted by 
many valleys of zero volume which intersect each other in a complicated 
way. The boundaries of those valleys have arbitrarily steep inclination
and the mountains are arbitrarily high. This is 
because the solutions of the zero volume condition are now of the 
form $j^{(v)}_1=j^{(v)}_0(j^{(v)}_2,..,j^{(v)}_{N_v})$ and we may let 
$j^{(v)}_2,..,j^{(v)}_{N_v}\to\infty$ in many 
ways (along many valleys). It is clear that now for each $v$ 
$V_v((j^{(v)}_1=j^{(v)}_0,j^{(v)}_2,..,j^{(v)}_{N_v}))-
V_v((j^{(v)}_1=j_0,j^{(v)}_2,..,j^{(v)}_{N_v})+\delta j^{(v)})$
with $\delta j^{(v)}_k=\delta_{kl},\;l\in\{1,..,N_v\}$ can become arbitrarily 
large. This is what we are going to prove explicitly in our companion 
paper \cite{16} based on the recent simplification of the analytical
expression for the matrix elements of the volume operator \cite{17}
\footnote{For the very special case of matrix elements we are going to use a similar expression was also derived in \cite{de Pietri} using a different method.}.

One may object that the diverging configurations 
$(j^{(v)}_1,j^{(v)}_2,..,j^{(v)}_{N_v})$ for each individual $v$ do not 
look isotropic in general and that one should require isotropy 
as in LQC which perhaps should enforce 
$j^{(v)}_1=..=j^{(v)}_{N_v}$ which could lead to a bound since the 
parameter space is now one dimensional\footnote{Indeed, if we choose 
$j_1=j_2=j_3=j$ in formula 
(\ref{e'(B)^2 Endresultat eichinvarianter 3-Vertex}) then the result is 
bounded in $j$.}. 
However, here we now invoke a second key difference 
between LQC and LQG: LQC is a quantum mechanical toy model of LQG. It can 
be thought of as LQG where all the vertices of a spin network state are to 
be thought of as equivalent due to homogeneity. In other words, LQC is 
like LQG except that we only consider spin network states with a single 
vertex (and a few edges, usually no more than three). LQG on the other 
hand is a field theory and spin network states 
can have an arbitrary number of vertices. Consider now a zero volume spin 
network 
state $T_{\lambda,v}$ where $\lambda$ is the eigenvalue of the inverse 
scale factor and $v$ is a vertex at which the graph $\gamma_v$ with 
a single non co -- planar\footnote{The volume operator vanishes on 
(non) gauge invariant 
vertices which are not at least (tri) four -- valent and at which not at 
least 
three tangents of the adjacent edges are linearly independent.} 
vertex $v$ of the spin network 
state is located. Let $S$ be a finite\footnote{This will be sufficient
if the spatial manifold is compact as for $k=1$ FRW models or if we are 
interested only a compact, observationally accessible (inside the Hubble 
radius) region. In the 
$k=0,-1$ case we can use in the infinite tensor product \cite{19}.} 
set of $n$ vertices obtained from a random 
sprinkling of points into the spatial manifold \cite{18} and consider the 
state $\psi:=\otimes_{v\in S} T_{\lambda,v}$. 
We may arrange that the $\gamma_v$ are mutually 
disjoint 
and that their orientation is random as well. This state is 
homogeneous and isotropic on large scales but not on microscopic ones 
(the violation scale of homogeneity and isotropy depends on the 
number of points in $S$ and decreases with $n$). The state is normalized 
and still an eigenvalue $\lambda$ eigenstate of the inverse scale factor.
This proves that we may construct kinematical states with large scale 
homogeneity and isotropy in LQG with arbitrarily large norm of the inverse 
scale factor\footnote{This is a purely kinematical result as in LQC.
In fact it must be because the inverse scale 
factor itself is not a Dirac observable, see next section.}.\\
\\
Let us now go into more details:\\
In order to match the LQC calculations as closely as possible, what we 
really have studied is the contribution of a scalar 
(or inflaton) field to the Hamiltonian constraint (smeared energy density)
operator whose classical expression is 
\be \label{3.1}
2C_{scalar}(N)=\int_\sigma\; d^3x\; N\;[\frac{\pi^2}{\sqrt{\det(q)}}+
\sqrt{\det(q)} (q^{ab} \phi_{,a}\phi_{,b}+2V(\phi))]
\ee
Here $\phi$ is the scalar field, $\pi$ its conjugate momentum,
$q_{ab}$ is the pull back to the spatial manifold $\sigma$ of the four 
metric $g_{\mu\nu}$ on the four manifold $M\cong \Rl\times \sigma$ and 
$N$ is a test function (see also section \ref{s0}). In an exactly 
homogeneous situation, say the FRW
metric, the derivative term drops out and the potential term becomes 
proportional to $a^3$ while the kinetic term becomes proportional to 
$a^{-3}$ where $a$ is the scale factor of the FRW metric. The 
classical initial singularity $\lim_{t\to 0} a(t)=0$ is therefore 
prominent in the kinetic term. Of course in full LQG the kinetic term must 
be and can be considered as well, see \cite{10b}, but it will be 
sufficient to consider the kinetic term to demonstrate our point.

Consider therefore the kinetic term
\be \label{3.2}
2C^{kin}_{scalar}(N)=\int_\sigma\; d^3x\; N\;\frac{\pi^2}{\sqrt{\det(q)}}
\ee
Its quantization is derived in \cite{10b} on the kinematical Hilbert
space ${\cal H}_{kin}={\cal H}_{geo}\otimes {\cal H}_{scalar}$.
An orthonormal basis in this Hilbert space \cite{10d} consists 
of the 
charge -- spin network states $T_{\gamma,\vec{j},\vec{I}}\otimes 
T_{V(\gamma),\vec{n}}$. Here $\gamma$ is a graph with vertex set 
$V(\gamma)$, $\vec{j}$ denotes a labelling of each edge $e$ of $\gamma$ by 
half integral spin quantum numbers $j_e$, and $\vec{I},\;\vec{n}$ 
respectively denote a labelling 
of each vertex $v$ of $\gamma$ with a gauge invariant intertwiner $I_v$ 
and integer $n_v$ respectively. The action of the corresponding 
operator, densely defined in the finite linear span of this basis, 
can now schematically be written as 
\be \label{3.3}
2\widehat{C^{kin}_{scalar}}(N)\;T_{\gamma,\vec{j},\vec{I},\vec{n}}=
\sum_{v\in V(\gamma)}\; N(v)\;[\widehat{\frac{1}{\sqrt{\det(q)}}(v)}\;
T_{\gamma,\vec{j},\vec{I}}]\;\otimes\;[\widehat{\pi}(v)^2\;
T_{V(\gamma),\vec{n}}]
\ee
The explicit meaning and action of the genuine operators ({\it not} 
operator valued distributions) can be found in \cite{10b,16a} and is 
also reviewed in \cite{16} for the convenience of the reader. For the 
purpose of this paper it is enough to report the result of our 
calculation. 

\begin{figure}[hbt]
    \label{fig1}
    \center
    \cmt{7}{
    \psfrag{v1}{${v_1}$}
    \psfrag{v2}{${v_2}$}
    \psfrag{e1}{$e_1$}
    \psfrag{e2}{$e_2$}
    \psfrag{e3}{$e_3$}
    \includegraphics[height=3cm]{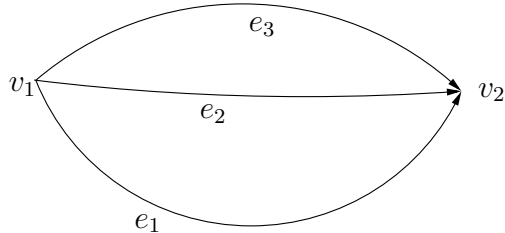} 
    \caption{Simplest 3 valent gauge invariant graph $\gamma$: $V(\gamma)=\{v_1,v_2\},E(\gamma)=\{e_1,e_2,e_3\}$} }
\end{figure}

The analogon of the LQC inverse scale factor (cubed) in LQG is of course 
the non -- negative operator $\widehat{\frac{1}{\sqrt{\det(q)}}(v)}$. 
To show that it is unbounded from above on zero volume eigenstates it 
is sufficient to construct one single system of states whose members 
have arbitrarily large norm. The simplest 
zero volume eigenstates are spin network states 
$T_{\gamma,\vec{j},\vec{I}}$ which have at most three valent vertices.
They are very simple because the intertwiners $I_v$ are then uniquely 
determined. We are thus interested in the numbers
\be \label{3.4}
||\widehat{\frac{1}{\sqrt{\det(q)}}(v)}\;T_{\gamma,\vec{j}}||^2
=<T_{\gamma,\vec{j}},[\widehat{\frac{1}{\sqrt{\det(q)}}(v)}]^2
\;T_{\gamma,\vec{j}}>
\ee
where $v$ is a three valent vertex of $\gamma$ whose adjacent, non co -- 
planar edges carry the spin quantum numbers $j_1 \le j_2 \le j_3$ subject 
to the constraint that $j_3\in \{j_1+j_2,j_1+ j_2-1,..,|j_1-j_2|\}$. To 
make things definite, consider a graph $\gamma$ with three edges and two 
three valent, non co -- planar vertices, see figure \ref{fig1}.\\
%

Remarkably the number (\ref{3.4}) can be computed analytically. This is 
possible due to a recent improvement in the technology for the computation 
of 
the matrix elements of the volume operator for full LQG \cite{17}.
The analytical formula for (\ref{3.4}) is given by the surprisingly 
compact expression\\
%
%
\fcmt{18}{\[ \begin{array}{llll}
     \\
    ||\widehat{\frac{1}{\sqrt{\det(q)}}(v)}\;T_{\gamma,\vec{j}}||^2
   &=&\multicolumn{2}{l}{\displaystyle\frac{C^2}{(\ell_P)^6}\cdot\Bigg[
    \displaystyle\frac{32\,}{9\,
    {\left( 1 + 2\,{j_1} \right) }^2\,{\left( 1 + 2\,{j_2} \right) }^2\,{\left( 1 + 2\,{j_3} \right) }^2}}
    
    \\\\
    &&\times&\big[ 108\,A_1 \,A_2 \,
    A_3
      
    \\
    &&&
   - 3\,{\left( 2\,{\left( -1 \right) }^{2\,{j_1}} + 
           {\left( -1 \right) }^{2\,{j_3}} \right) }^2\,A_2 \,
       {\left( A_1 -A_2 + A_3 \right) }^2 
      
    \\
    &&& 
       
       -3\,{\left( 2\,{\left( -1 \right) }^{2\,{j_2}} + {\left( -1 \right) }^{2\,{j_3}} \right) }^2\,
       
       A_1  \,{\left( -A_1 + A_2 + A_3 \right) }^2
       
    \\
    &&&
      
       -3\,{\left( 1 + 2\,{\left( -1 \right) }^{2\,\left( {j_1} + {j_2} \right) } \right) }^2\,
       A_3 \,{\left( A_1 + A_2 - 
           A_3  \right) }^2 \ 
    \\
    &&&  
       -~ \left( 1 + 2\,{\left( -1 \right) }^{2\,\left( {j_1} + {j_2} \right) } \right) \,
       \left( 2\,{\left( -1 \right) }^{2\,{j_1}} + {\left( -1 \right) }^{2\,{j_3}} \right) \,
       \left( 2\,{\left( -1 \right) }^{2\,{j_2}} + {\left( -1 \right) }^{2\,{j_3}} \right) \,
    \\
    &&&~~~~~~~\times
       \left( -A_1 +A_2 +A_3 \right) \,
       \left( A_1 -A_2 + A_3 \right) \,
       \left( A_1 + A_2 - A_3  \right)  
       
     \big]  
     \\\\&&\times&	   
     \,
    {\left( {V_{1A}}^{\frac{1}{4}} - {V_{1B}}^{\frac{1}{4}} \right) }^2\,
    {\left( {V_{2A}}^{\frac{1}{4}} - {V_{2B}}^{\frac{1}{4}} \right) }^2\,
    {\left( {V_{3A}}^{\frac{1}{4}} - {V_{3B}}^{\frac{1}{4}} \right) }^2\Bigg]^2

\end{array}\] \be\label{e'(B)^2 Endresultat eichinvarianter 3-Vertex} \ee }

~\\\\
where $C$ is a positive  numerical constant depending on the regularization of the operator 
$\widehat{\frac{1}{\sqrt{\det(q)}}(v)}$ and the volume operator.
Furthermore $(j_1,j_2,j_3)$ are the spins of the 3 edges $(e_1,e_2,e_3)$ outgoing at the vertex $v$ and $A_K:=j_K(j_K+1)$. Moreover

\[\begin{array}{lcl}
    V_{1A}
    &=&\big[(-j_1+j_2+j_3+1)(j_1-j_2+j_3)(j_1+j_2-j_3)(j_1+j_2+j_3+1) \big]^{\frac{1}{2}}
    
    \\
    V_{1B}
    &=&\big[(-j_1+j_2+j_3)(j_1-j_2+j_3+1)(j_1+j_2-j_3+1)(j_1+j_2+j_3+2) \big]^{\frac{1}{2}}
			    
    \\
    
    V_{2A}
    &=&\big[(-j_1+j_2+j_3)(j_1-j_2+j_3+1)(j_1+j_2-j_3)(j_1+j_2+j_3+1) \big]^{\frac{1}{2}}

    \\
   
   V_{2B}
       &=&\big[(-j_1+j_2+j_3)(j_1-j_2+j_3)(j_1+j_2-j_3+1)(j_1+j_2+j_3+2) \big]^{\frac{1}{2}}

    \\
    
    V_{3A}
        &=&\big[(-j_1+j_2+j_3)(j_1-j_2+j_3)(j_1+j_2-j_3+1)(j_1+j_2+j_3+1) \big]^{\frac{1}{2}}
			    
    \\			    
    V_{3B}
        &=&\big[(-j_1+j_2+j_3+1)(j_1-j_2+j_3+1)(j_1+j_2-j_3)(j_1+j_2+j_3+2) \big]^{\frac{1}{2}}

\end{array}\]

We now display a few diverging series of configurations $(j_1,j_2,j_3)$ where 
we  always plot \linebreak
$||\widehat{\frac{1}{\sqrt{\det(q)}}(v)}\;T_{\gamma,\vec{j}}||:=\sqrt{||\widehat{\frac{1}{\sqrt{\det(q)}}(v)}\;T_{\gamma,\vec{j}}||^2}$, neglecting the prefactor $\frac{C}{(\ell_P)^3)}$\\
\pagebreak
\subsubsection*{''Oscillating''}
\paragraph*{\fbox{$j_1=j_2=\frac{j_3}{2}$}}%
If we set $j_1=j_2=\frac{j_3}{2}$ where $j_3 \in \mb{N}$, we get:

\begin{footnotesize}
\be
||\widehat{\frac{1}{\sqrt{\det(q)}}(v)}\;T_{\gamma,\vec{j}}||
   \propto
\frac{128\,{\sqrt{2}}\,\left( -1 + {\left( -1 \right) }^{{j_3}} \right) \,{{j_3}}^4\,
    \left( -3\,\left( 2 + {\left( -1 \right) }^{{j_3}} \right)  + 
      \left( 1 + {\left( -1 \right) }^{3\,{j_3}} \right) \,{j_3} \right) }{9\,\left( 1 + {j_3} \right) \,
    {\left( 1 + 2\,{j_3} \right) }^{\frac{7}{4}}}
\ee

\begin{figure}[htbp!]        
  \begin{minipage}[t]{7.5cm}
  \psfrag{e2}{\tiny\hspace{-5mm}$\small\|\widehat{\frac{1}{\sqrt{\det(q)}}(v)}\;T_{\gamma,\vec{j}}\|$}
  \psfrag{j3}{$j_3$}
  \includegraphics[height=4.5cm]{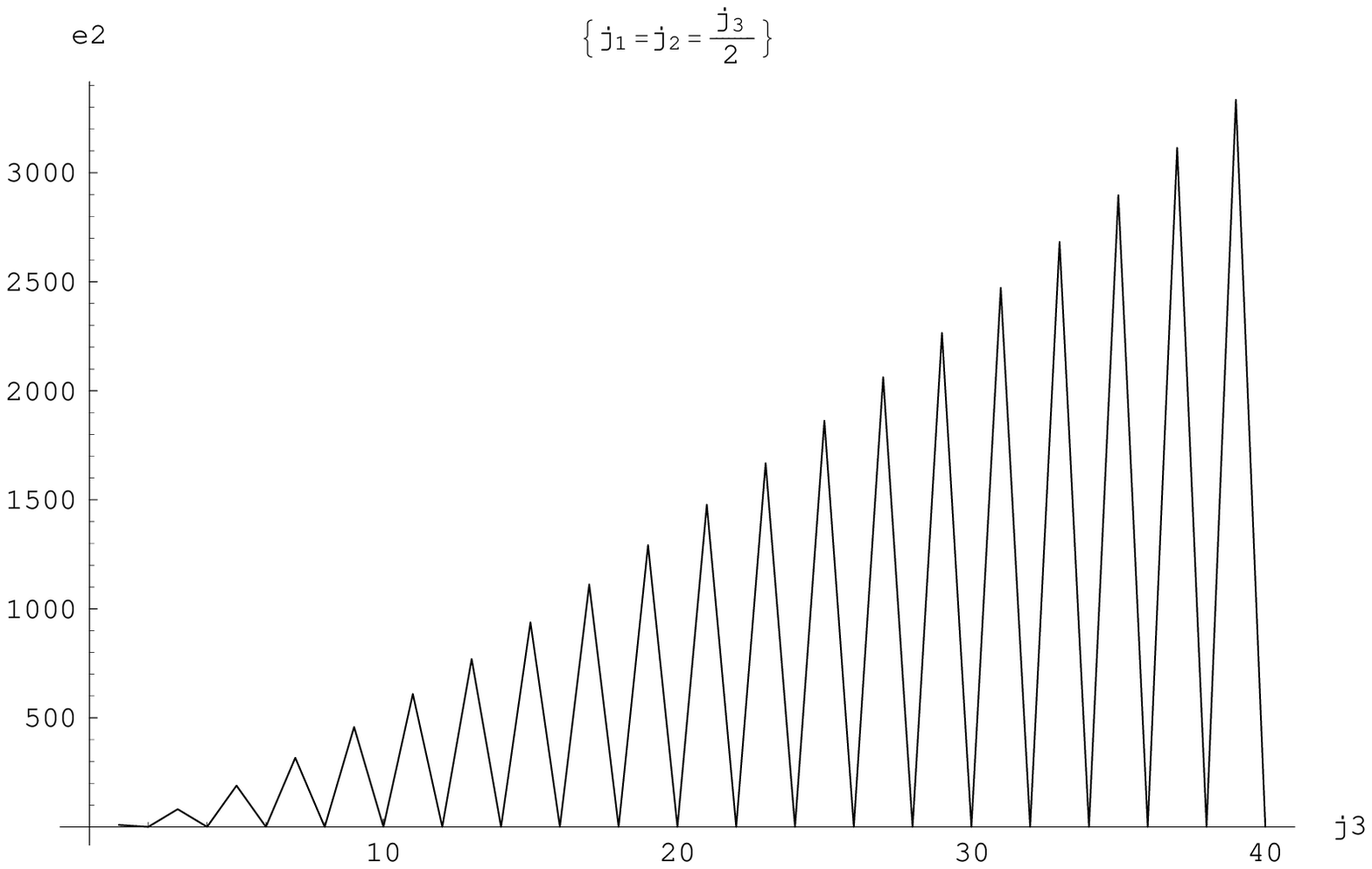}
  \caption{Plot for $j_1=j_2=\frac{j_3}{2}$ where $j_3 \in \mb{N}$ with $1\le j_3 \le 40$. The graph oscillates between 0 (if $j_3$ even) and an increasing value (if $j_3$ odd)} 
  \end{minipage}
 \begin{minipage}[b]{8cm}
 Asymtotically this increases as
  \[\begin{array}{lcl}
     ||\widehat{\frac{1}{\sqrt{\det(q)}}(v)}\;T_{\gamma,\vec{j}}||
    & \propto&
    \frac{128\cdot 6 \cdot\sqrt{2} \cdot {j_3}^4}{9(1+j_3)(1+2j_3)^{\frac{11}{4}}}\nonumber\\
    &\stackrel{j_3 \rightarrow \infty}{\propto}&
    35.9 \cdot {j_3}^{\frac{5}{4}}
  \end{array}\]
  \vfill
  \end{minipage}
\end{figure}

\end{footnotesize}

\paragraph*{\fbox{$j_1=j_2=\frac{j_3}{2}+\frac{1}{2}$}}
\begin{samepage}

If we set $j_1=j_2=\frac{j_3}{2}+\frac{1}{2}$ where $j_3 \in \mb{N}$, we get:

\begin{footnotesize}
\be\begin{array}{lcl}
||\widehat{\frac{1}{\sqrt{\det(q)}}(v)}\;T_{\gamma,\vec{j}}||
   &\propto&
    \frac{1}{9\,
    {\left( 2 + {j_3} \right) }^4\,{\left( 1 + 2\,{j_3} \right) }^2}
    
   \\\\
   &&\times~    
    128\,{\sqrt{2}}\,\left( 1 + {\left( -1 \right) }^{{j_3}} \right) \,
    {{j_3}}^2\,{\left( 1 + {j_3} \right) }^3\,\left( -3\,
       \left( -7 + 3\,{\left( -1 \right) }^{{j_3}} + {\left( -1 \right) }^{3\,{j_3}} \right)  + 
      \left( -1 + {\left( -1 \right) }^{3\,{j_3}} \right) \,{j_3} \right) \,
    \\
    &&\times
    {\left( -\left( 2^{\frac{1}{4}}\,{\left( {{j_3}}^2\,\left( 1 + {j_3} \right)  \right) }^{\frac{1}{8}} \right)
            + {\left( {\left( 1 + {j_3} \right) }^2\,\left( 3 + 2\,{j_3} \right)  \right) }^{\frac{1}{8}} \right)
        }^2\,
	
    \\&&
    \times{\left( {\left( {j_3}\,{\left( 1 + {j_3} \right) }^2 \right) }^{\frac{1}{8}} - 
        {\left( {j_3}\,\left( 3 + 5\,{j_3} + 2\,{{j_3}}^2 \right)  \right) }^{\frac{1}{8}} \right) }^4	
\end{array}\ee
\vspace{-1cm}
\begin{figure}[hbt]
         \begin{minipage}[t]{7.5cm}
	 \psfrag{e2}{\tiny\hspace{-5mm}$\small\|\widehat{\frac{1}{\sqrt{\det(q)}}(v)}\;T_{\gamma,\vec{j}}\|$}
	 \psfrag{j3}{$j_3$}
	 \includegraphics[height=4.5cm]{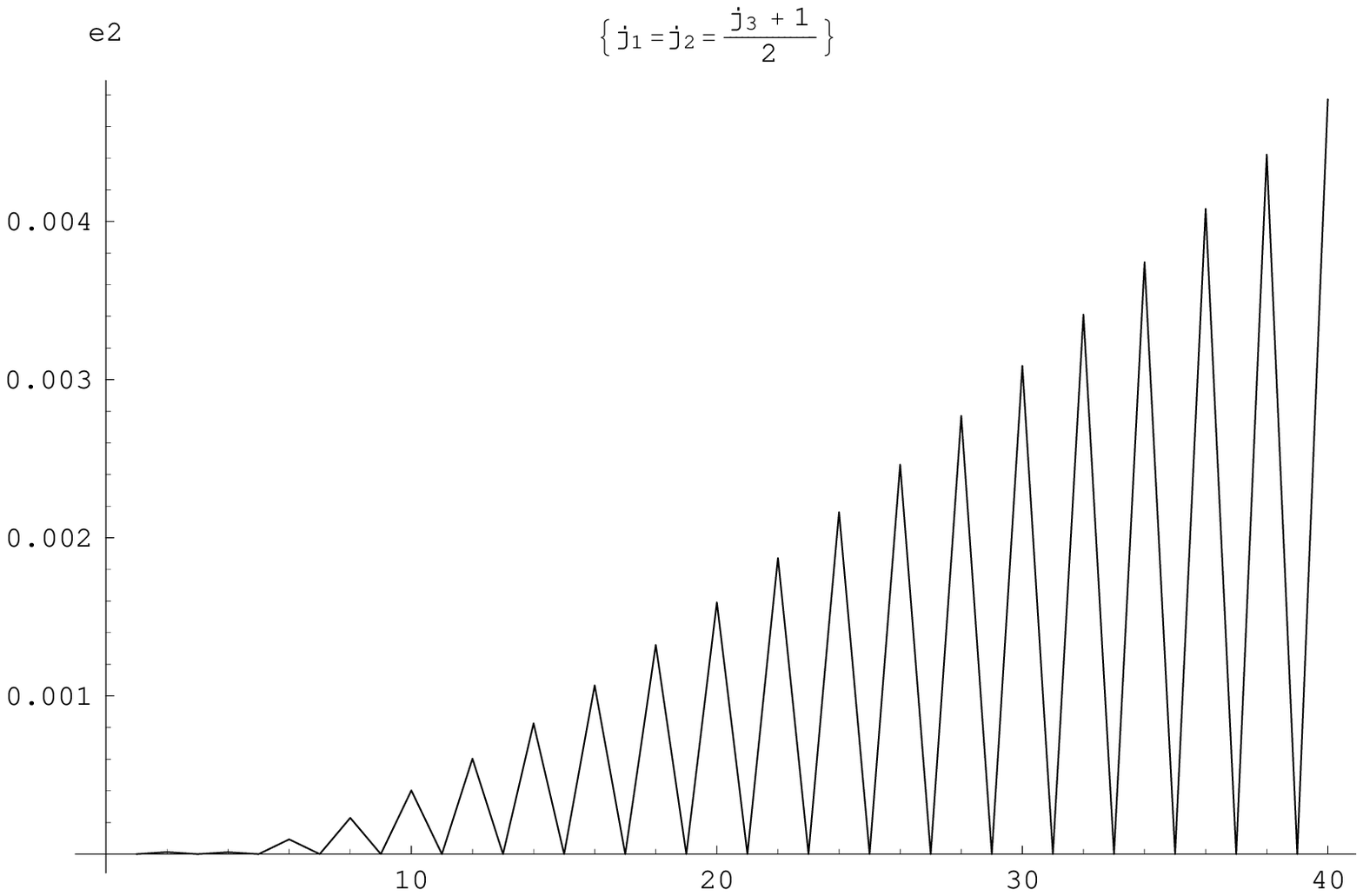}
          \caption{Plot for $j_1=j_2=\frac{j_3+1}{2}$ where $j_3 \in \mb{N}$ with $1\le j_3 \le 40$. The graph oscillates between 0 (if $j_3$ odd) and an increasing value (if $j_3$ even)} 
         \end{minipage}
         \begin{minipage}[b]{8cm}
            Asymtotically this increases as
            \[\begin{array}{lcl}
                ||\widehat{\frac{1}{\sqrt{\det(q)}}(v)}\;T_{\gamma,\vec{j}}||
                & \propto&5.9\cdot 10^{-5}~{j_3}^{\frac{5}{4}}
            \end{array}\] 
         \end{minipage}
\end{figure}   

\end{footnotesize}   
\end{samepage}

\subsubsection*{Increasing}
\paragraph*{\fbox{$j_1=\frac{3}{2}~ j_2=j_3+\frac{1}{2}$}}%
If we set $j_1=\frac{3}{2}$, $j_2=j_3+\frac{1}{2}$ where $j_3 \in \mb{N}$, we get:
\begin{footnotesize}
\[\begin{array}{lcl}
||\widehat{\frac{1}{\sqrt{\det(q)}}(v)}\;T_{\gamma,\vec{j}}||
   &\propto&\displaystyle

   	\frac{1}{3\,\left( 1 + {j_3} \right) \,
    {\left( 1 + 2\,{j_3} \right) }^2}
   \\&&\times
   4\,{j_3}\,\left( -9 + 21\,{j_3} + 14\,{{j_3}}^2 \right) \,
    {\left( -\left( 3^{\frac{1}{8}}\,{\left( {j_3}\,\left( 2 + {j_3} \right)  \right) }^{\frac{1}{8}} \right)
            + {\left( -3 + 4\,{j_3} + 4\,{{j_3}}^2 \right) }^{\frac{1}{8}} \right) }^2
    \\&&\times\,
    {\left( -2\,{\left( {j_3}\,\left( 2 + {j_3} \right)  \right) }^{\frac{1}{8}} + 
        {\sqrt{2}}\,3^{\frac{1}{8}}\,{\left( -3 + 4\,{j_3} + 4\,{{j_3}}^2 \right) }^{\frac{1}{8}} \right) }^2\,
    {\left( {\left( {j_3}\,\left( 3 + 2\,{j_3} \right)  \right) }^{\frac{1}{8}} - 
        {\left( -6 + 9\,{j_3} + 6\,{{j_3}}^2 \right) }^{\frac{1}{8}} \right) }^2

\end{array}\]

\begin{figure}[htbp!]
         \begin{minipage}[t]{7.5cm}
	    \psfrag{e2}{\tiny\hspace{-5mm}$\small\|\widehat{\frac{1}{\sqrt{\det(q)}}(v)}\;T_{\gamma,\vec{j}}\|$}
	    \psfrag{j3}{$j_3$}	 
	    \includegraphics[height=5cm]{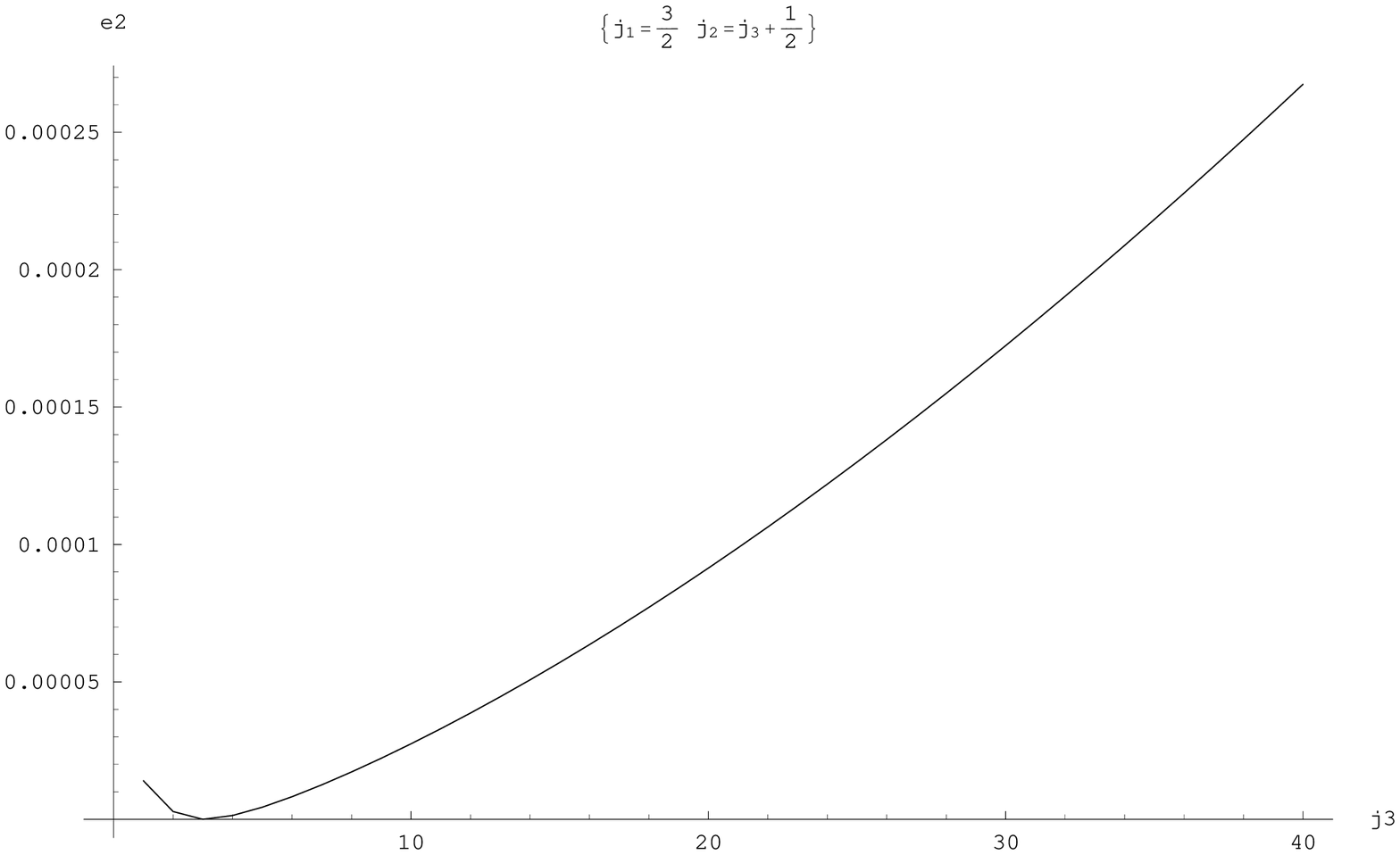}
              \caption{Plot for $j_1=\frac{3}{2}$, $j_2=j_3+\frac{1}{2}$ 
 where $j_3 \in \mb{N}$ with $1\le j_3 \le 40$. The graph first decreases 
for $1\le j_3 <3 $ and is 0 for $j_3=3$ . It increases for $j_3>3$}
	  \end{minipage}  
          \begin{minipage}[b]{8cm}
             Asymtotically this increases as
             \[\begin{array}{lcl}
                 ||\widehat{\frac{1}{\sqrt{\det(q)}}(v)}\;T_{\gamma,\vec{j}}||
                 & \propto& 1.06\cdot 10^{-6}~{j_3}^{\frac{3}{2}}
             \end{array}\]
          \end{minipage}
\end{figure}
\end{footnotesize}

%

\subsubsection*{\label{Nullkonfigurationen}Identical 0}
If we set $j_1=j_2=j_3$ and more general $j_1,j_2,j_3\in \mb{N}$ (all spins integer numbers) in (\ref{e'(B)^2 Endresultat eichinvarianter 3-Vertex})  then 

\[\begin{array}{lcl}
||\widehat{\frac{1}{\sqrt{\det(q)}}(v)}\;T_{\gamma,\vec{j}}||
   &=&0

\end{array}\] 
\vfill

\subsection*{General Configurations}

Finally we plot the numbers (\ref{3.4}) as $j_1,j_2$ vary over their 
allowed values at given $j_3$ for various values of $j_3$. The landscape of low valleys together with their arbitrarily high and steep boundaries is nicely illustrated:\\

Taking the general result (\ref{e'(B)^2 Endresultat eichinvarianter 3-Vertex}) (without the prefactors $(\ell_P)^9|Z|^\frac{3}{2}$)  we use the quantity

\[
  Q=\left\{\begin{array}{lcl}
                             30+\ln{[||\widehat{\frac{1}{\sqrt{\det(q)}}(v)}\;T_{\gamma,\vec{j}}||
			     (j_1,j_2,j_3)]}&&|j_1-j_2|\le j_3 \le j_1+j_2~~and~~j_1+j_2+j_3~is~integer~~\\
			     &&and~~ ||\widehat{\frac{1}{\sqrt{\det(q)}}(v)}\;T_{\gamma,\vec{j}}||(j_1,j_2,j_3)\ne 0
			     \\\\
			     0 &~~~&else
                         \end{array}\right.
\]
where the 30 is arbitrarily added in order to shift the plot upwards.
We make a three dimensional plot (in the range $\frac{1}{2}\le j_1 \le j_{max}$, $\frac{1}{2}\le j_2 \le j_{max}$ for each fixed value $5\le j_3 \le \frac{15}{2}$ :

I turns out that the non-zero configurations are grouped symetrically along lines parallel to the $j_1=j_2$ axis. The reason for this is of course the integer requirement $j_1+j_2+j_3\stackrel{!}{=}integer$.
Therefore we will get contributions on the $j_1=j_2$-axis only if $j_3$ is integer.
Because (\ref{e'(B)^2 Endresultat eichinvarianter 3-Vertex}) is symmetric with respect to the interchange of $j_1 \leftrightarrow j_2$ we may restrict  ourselves to the range $j_1\ge j_2$.

Additionally we contribute for each of those 3D-plot a 2 dimensional plot along the lines

\[
  j_2=j_1 - l~~~~~~\mbox{mit}~~0\le l\le \min[j_3,j_{max}-\frac{1}{2}]~,~~~ l+\frac{1}{2}\le j_1 \le j_{max}=25
\]

The restriction for the parameter $l$ is a result from the requirements $|j_1-j_2|\le j_3 \le j_1+j_2$ from which for $j_1>j_2$ we may remove the modulus. 
 In order to give a better impression we have joined only non-vanishing 
 values of $Q$ along the lines described above\footnote{For integer $j_3$ 
 also every second configuration on the lines gives a 0 (the identically
$0$ case). Hence, if we would join all the points, the plots would 
 oscillate between 0 and the now plotted curves and it would be hardly 
possible to 
see anything useful from them.}.

 \clearpage
 \begin{figure}[htbp!]
         \cmt{8}{
	 \psfrag{j_1}{$j_1$}
	 \psfrag{j_2}{$j_2$}
	 \psfrag{ln[eB^2]}{$Q$}
	 \psfrag{j_3=5}{}
	 \includegraphics[height=6cm]{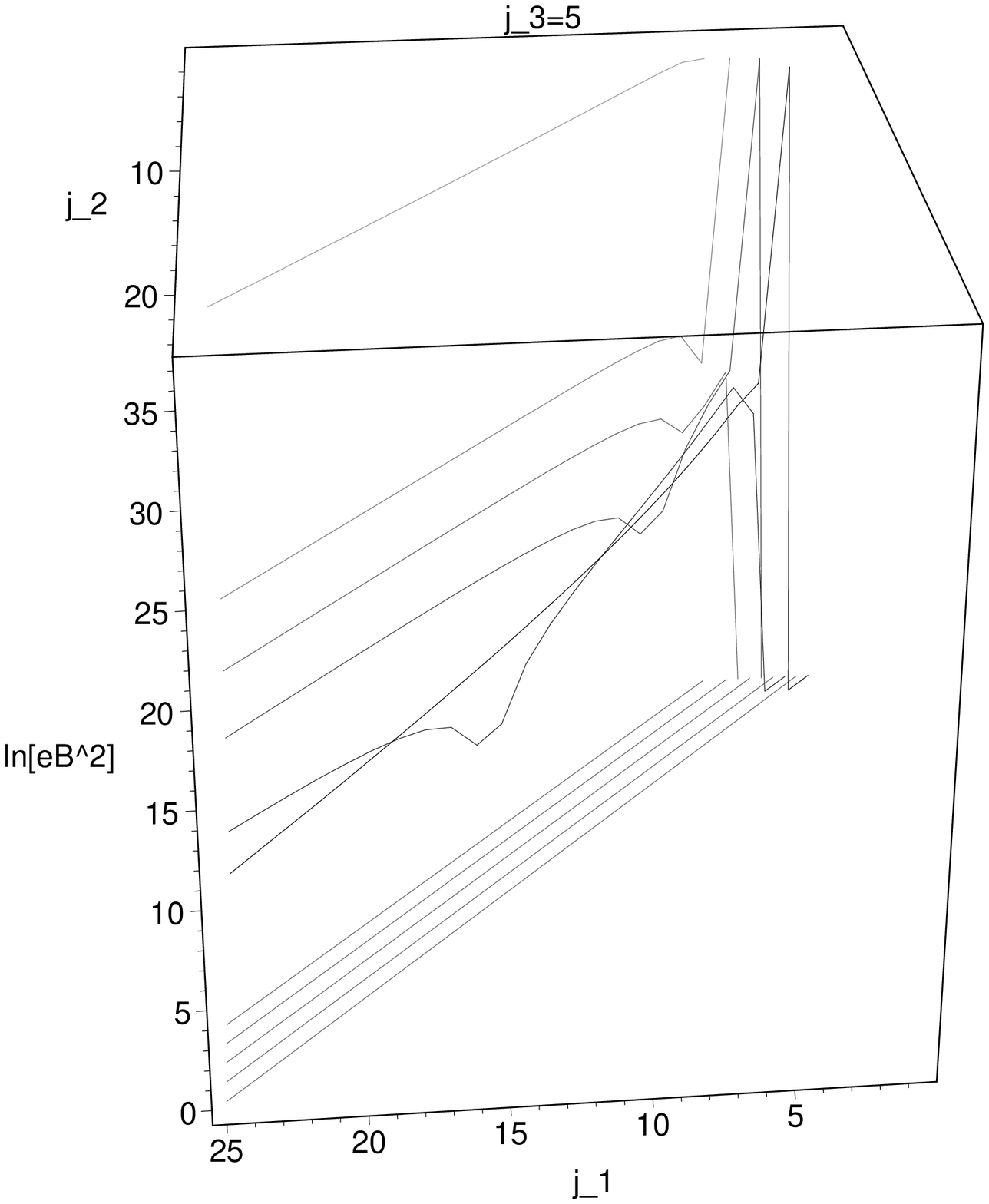}
          \caption{Plot for $j_3=5$} }
	 \cmt{8}{
	 \psfrag{j_1}{$j_1$}
	 \psfrag{j_2}{$j_2$}
	 \psfrag{ln[eB^2]}{$Q$}
	 \psfrag{j_3=5}{}
	 
	 \includegraphics[height=6cm]{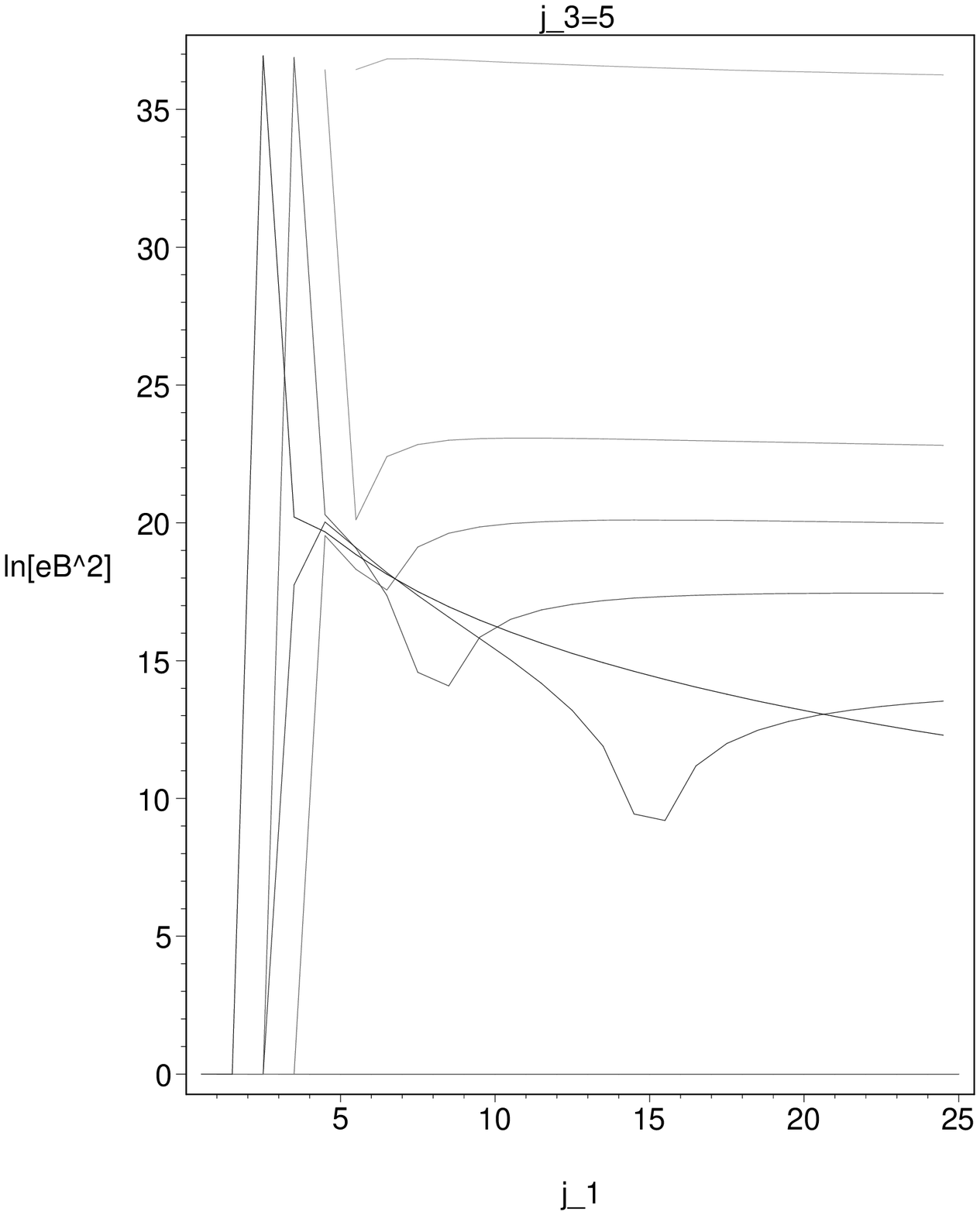}
          \caption{Plot for $j_3=5$} }
 
\end{figure}

\begin{figure}[htbp!]
         \cmt{8}{
	 \psfrag{j_1}{$j_1$}
	 \psfrag{j_2}{$j_2$}
	 \psfrag{ln[eB^2]}{$Q$}
	 \psfrag{j_3=11/2}{}
	 
	 \includegraphics[height=6cm]{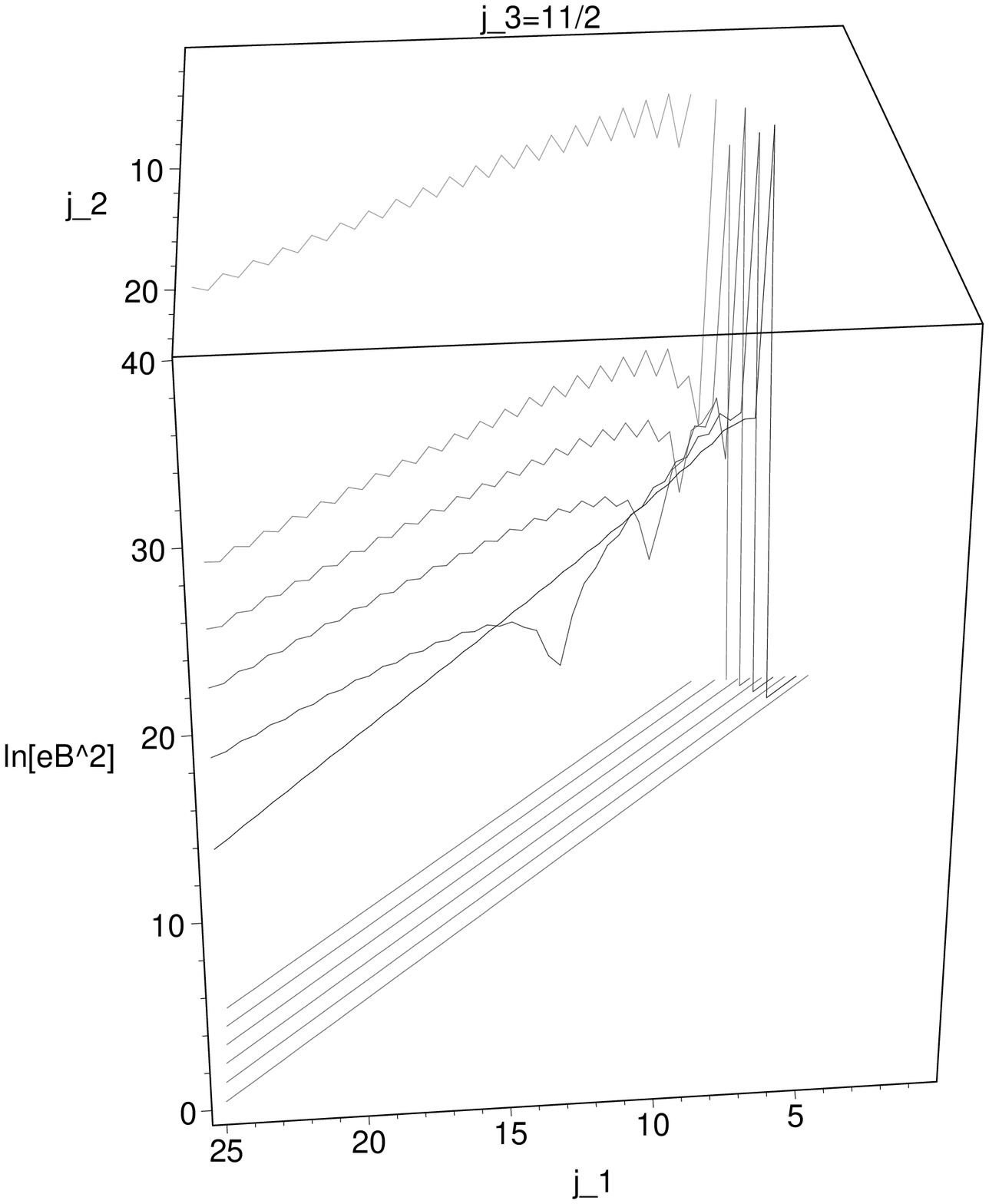}
          \caption{Plot for $j_3=\frac{11}{2}$} }
	 \cmt{8}{
	 \psfrag{j_1}{$j_1$}
	 \psfrag{j_2}{$j_2$}
	 \psfrag{ln[eB^2]}{$Q$}
	 \psfrag{j_3=11/2}{}
	 \includegraphics[height=6cm]{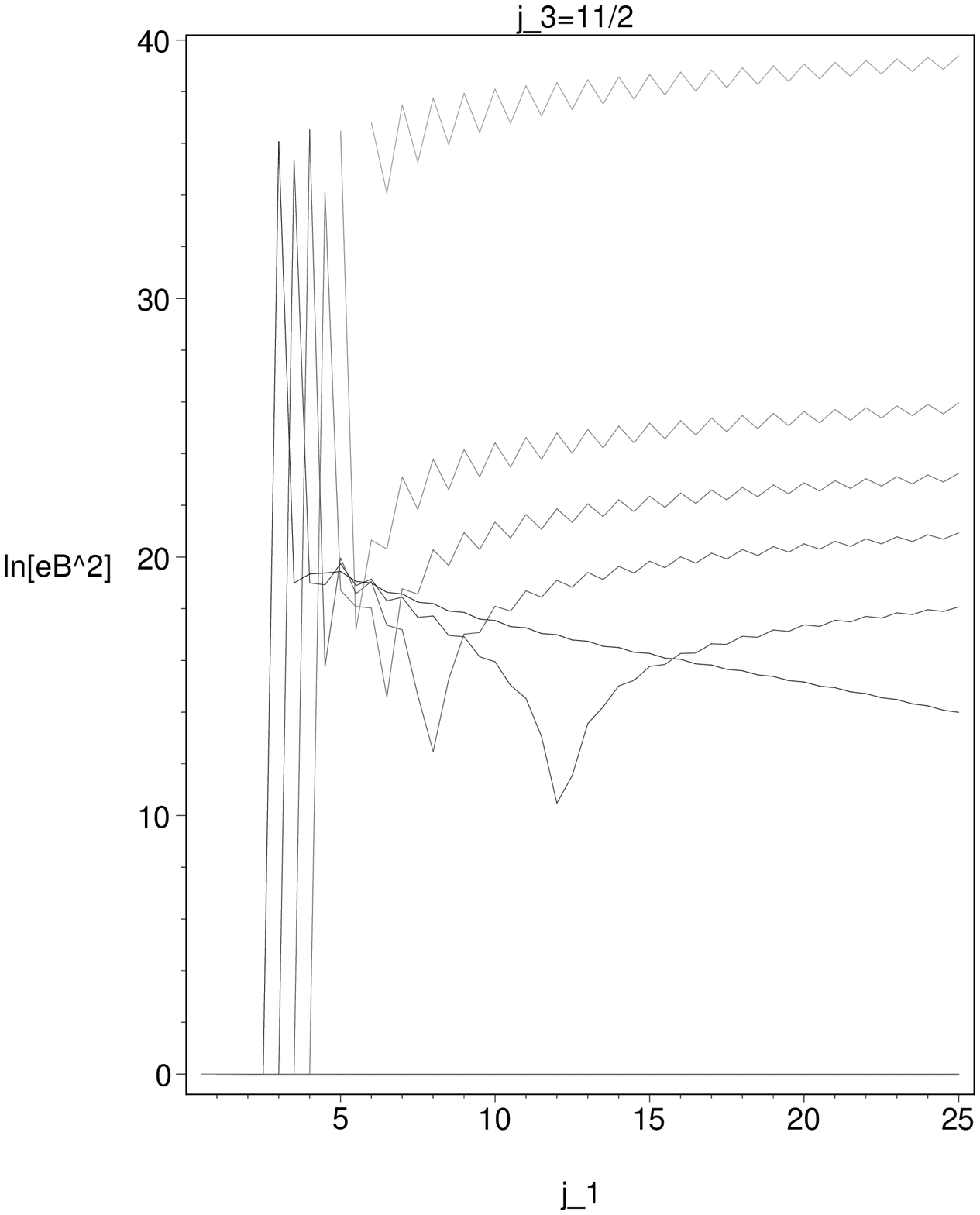}
          \caption{Plot for $j_3=\frac{11}{2}$, $j_2=j_1-l$. The different curves are (bottom to top):
	  $l=\frac{1}{2},\frac{3}{2},\ldots,\frac{11}{2}$} }
 
\end{figure}

\begin{figure}[htbp!]
         \cmt{8}{
	 \psfrag{j_1}{$j_1$}
	 \psfrag{j_2}{$j_2$}
	 \psfrag{ln[eB^2]}{$Q$}
	 \psfrag{j_3=6}{}
	 \includegraphics[height=6cm]{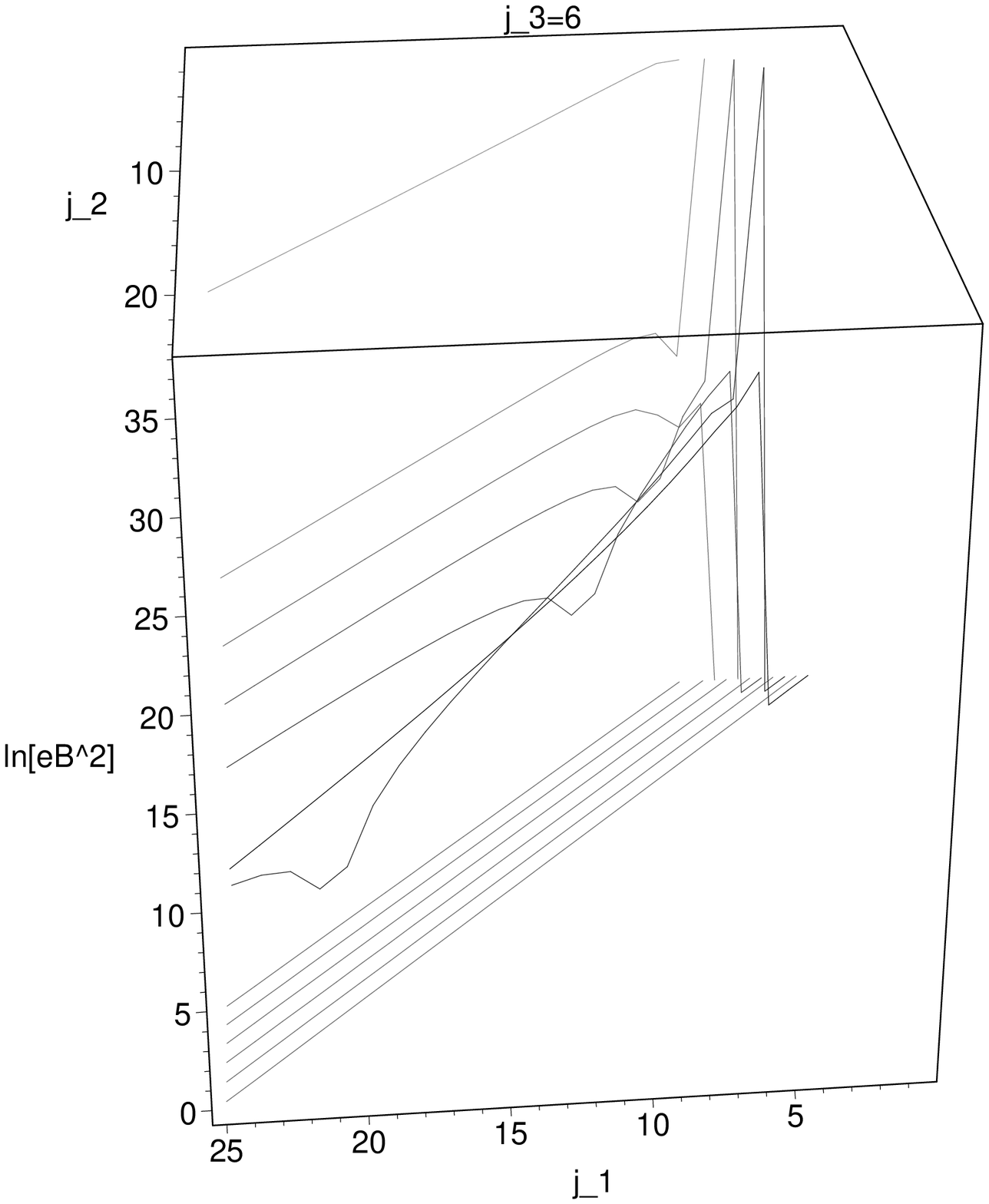}
          \caption{Plot for $j_3=6$} }
	 \cmt{8}{
	 \psfrag{j_1}{$j_1$}
	 \psfrag{j_2}{$j_2$}
	 \psfrag{ln[eB^2]}{$Q$}
	 \psfrag{j_3=6}{}
	 \includegraphics[height=6cm]{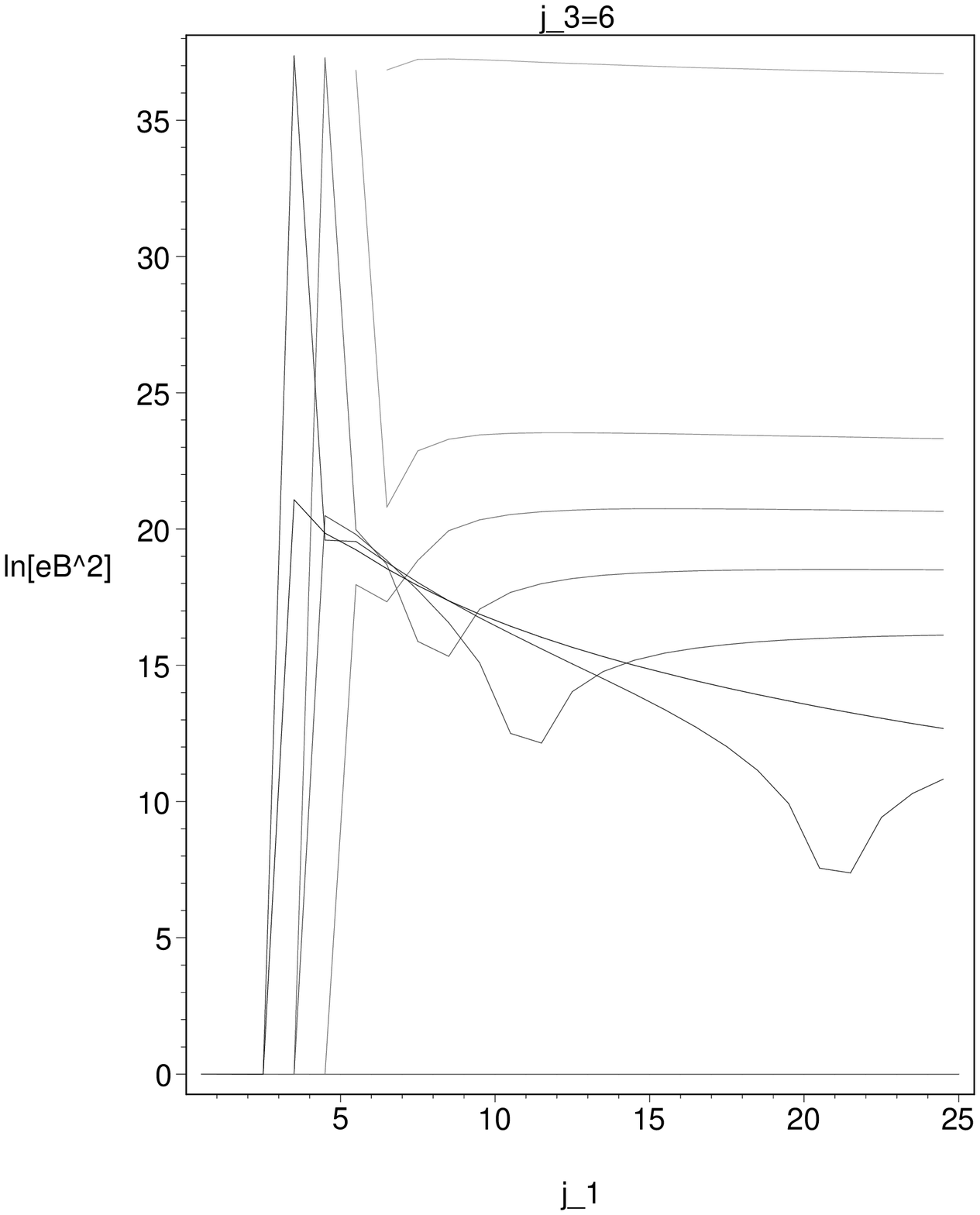}
          \caption{Plot for $j_3=6$} }
 
\end{figure}

\begin{figure}[htbp!]
         \cmt{8}{
	 \psfrag{j_1}{$j_1$}
	 \psfrag{j_2}{$j_2$}
	 \psfrag{ln[eB^2]}{$Q$}
	 \psfrag{j_3=13/2}{}
	 \includegraphics[height=6cm]{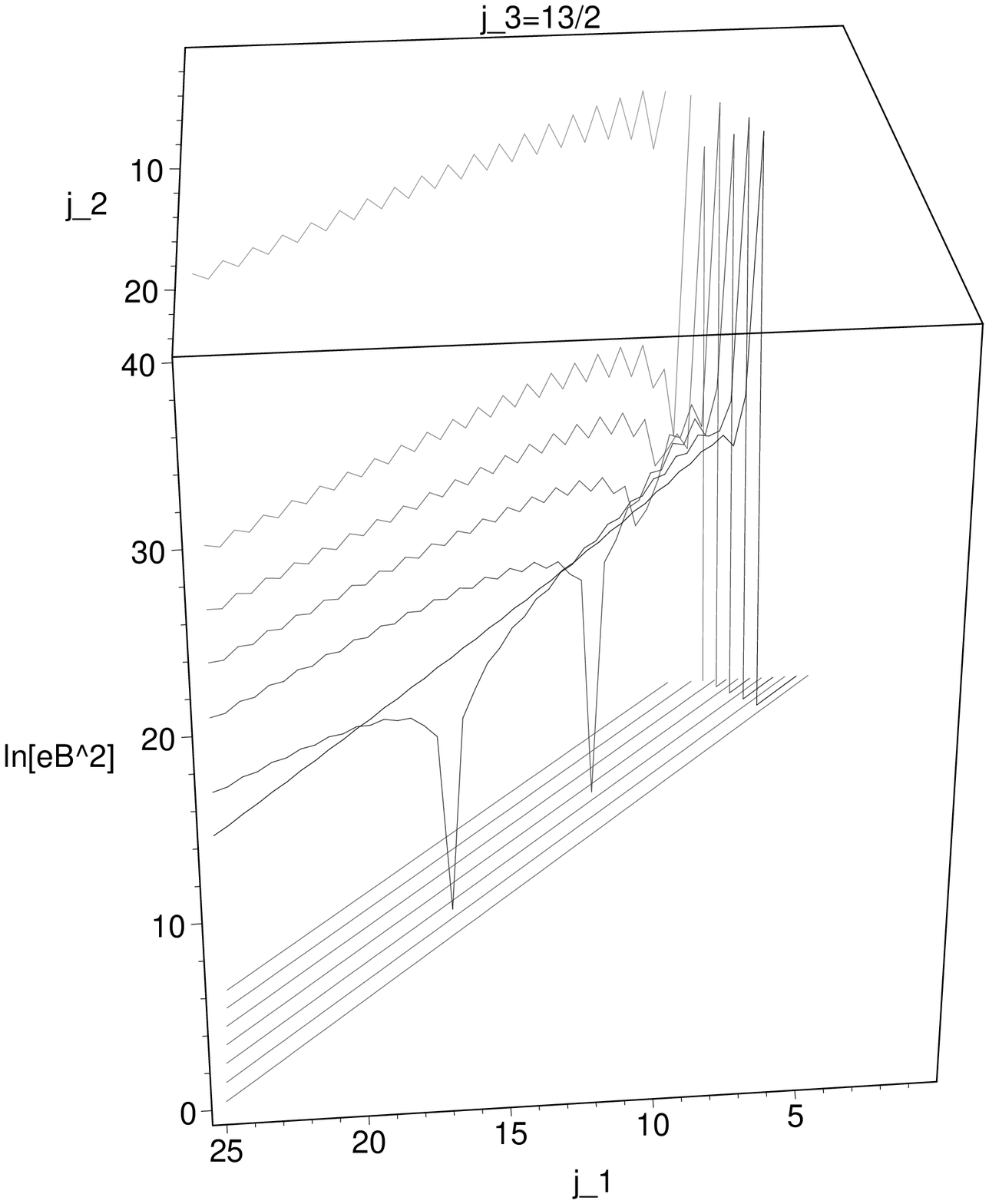}
          \caption{Plot for $j_3=\frac{13}{2}$} }
	 \cmt{8}{
	 \psfrag{j_1}{$j_1$}
	 \psfrag{j_2}{$j_2$}
	 \psfrag{ln[eB^2]}{$Q$}
	 \psfrag{j_3=13/2}{}
	 \includegraphics[height=6cm]{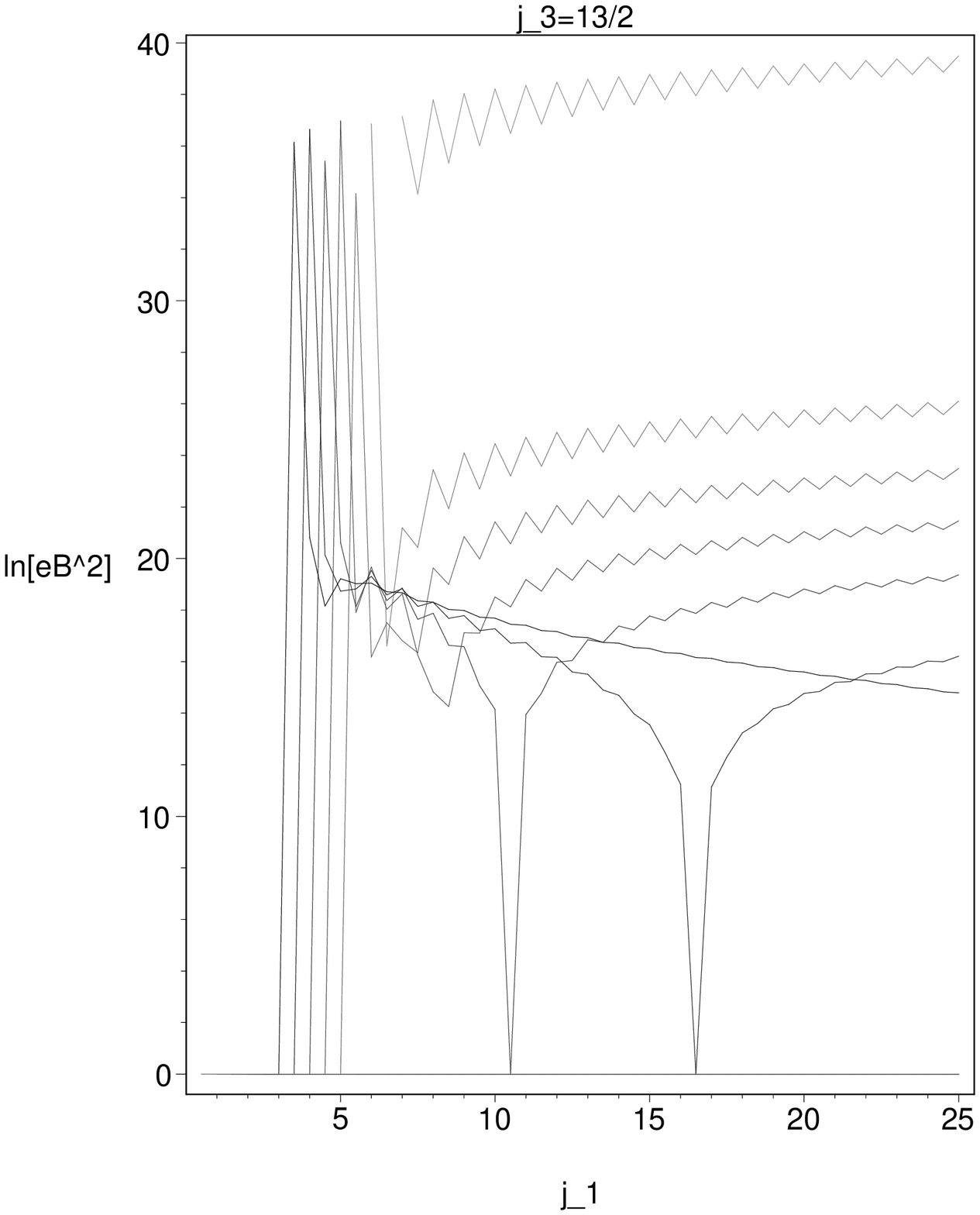}
          \caption{Plot for $j_3=\frac{13}{2}$, $j_2=j_1-l$. The different curves are (bottom to top):
	  $l=\frac{1}{2},\frac{3}{2},\ldots,\frac{13}{2}$} }
 \end{figure}

\begin{figure}[htbp!]
         \cmt{8}{
	 \psfrag{j_1}{$j_1$}
	 \psfrag{j_2}{$j_2$}
	 \psfrag{ln[eB^2]}{$Q$}
	 \psfrag{j_3=7}{}
	 \includegraphics[height=6cm]{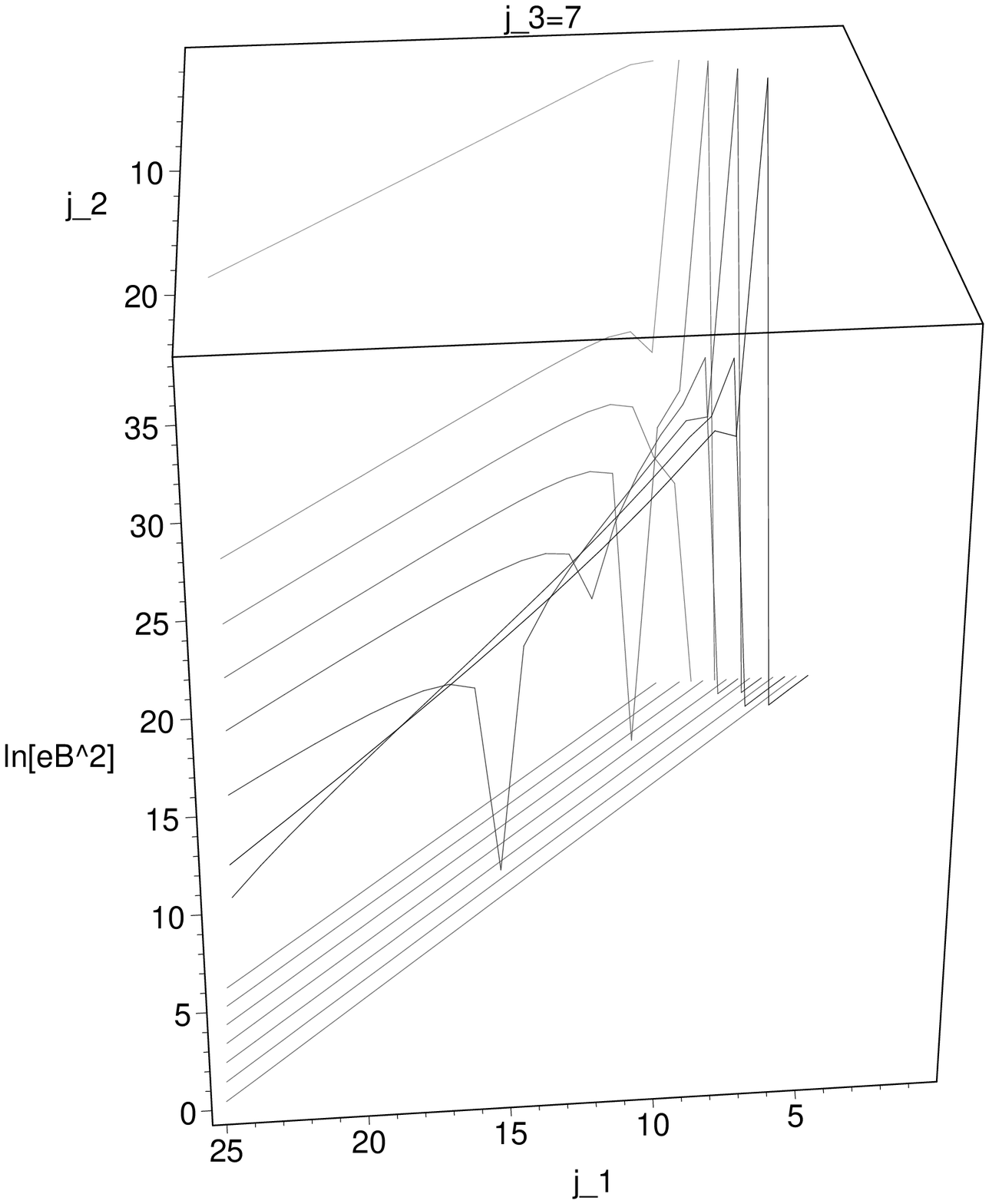}
          \caption{Plot for $j_3=7$} }
	 \cmt{8}{
	 \psfrag{j_1}{$j_1$}
	 \psfrag{j_2}{$j_2$}
	 \psfrag{ln[eB^2]}{$Q$}
	 \psfrag{j_3=7}{}
	 \includegraphics[height=6cm]{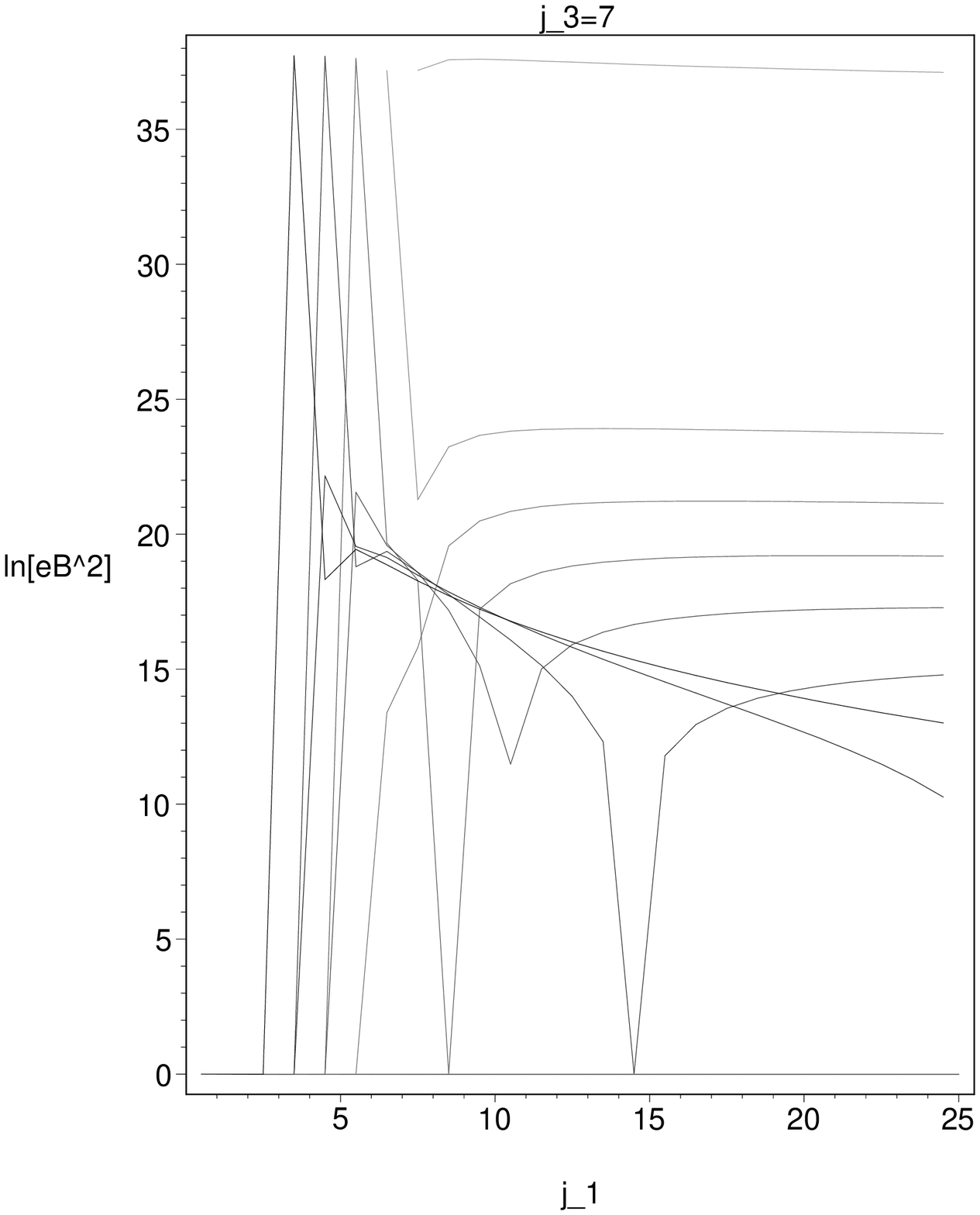}
          \caption{Plot for $j_3=7$} }	   
\end{figure}

\begin{figure}[htbp!]
         \cmt{8}{
	 \psfrag{j_1}{$j_1$}
	 \psfrag{j_2}{$j_2$}
	 \psfrag{ln[eB^2]}{$Q$}
	 \psfrag{j_3=15/2}{}
	 \includegraphics[height=6cm]{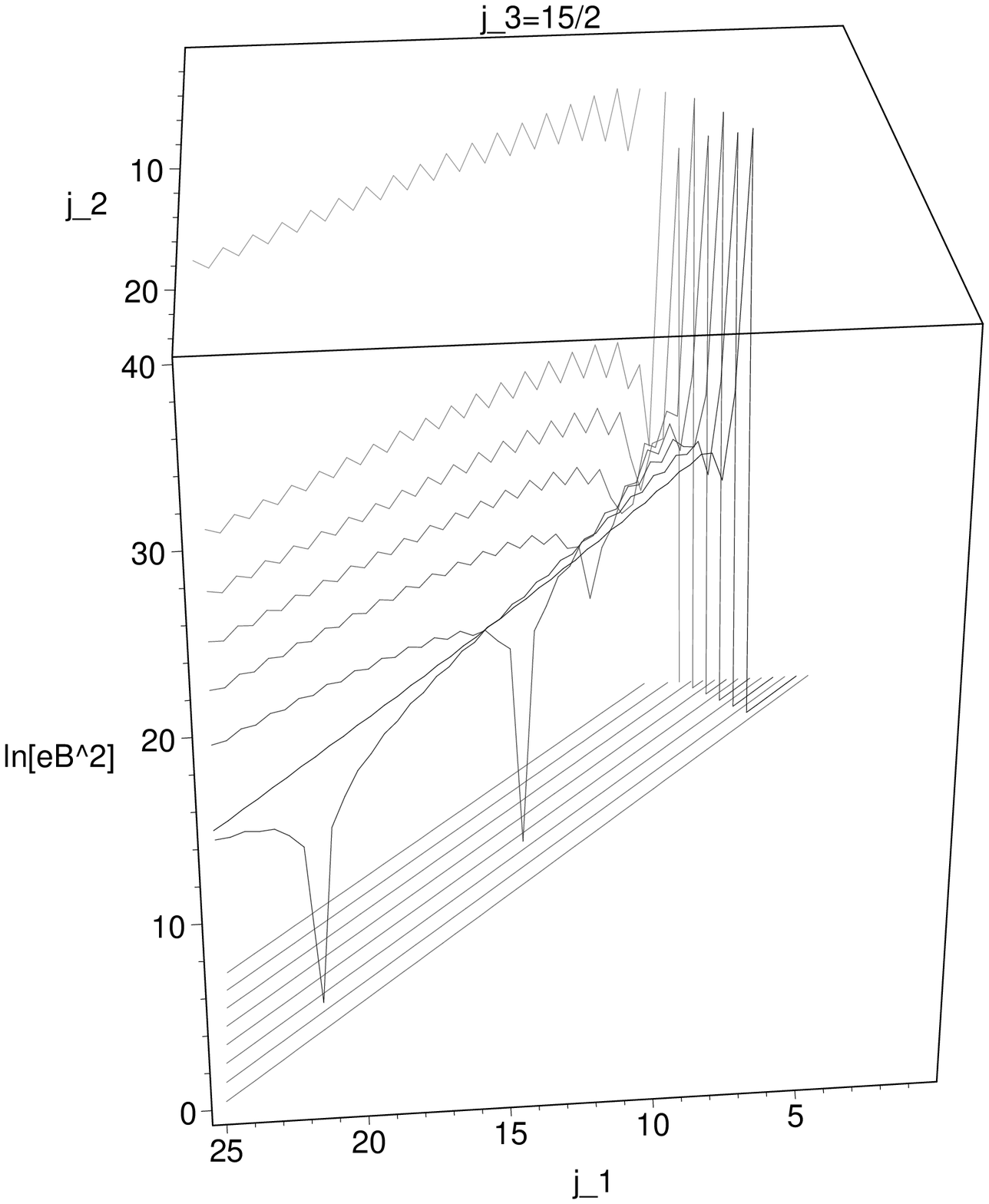}
          \caption{Plot for $j_3=\frac{15}{2}$} }
	 \cmt{8}{
	 \psfrag{j_1}{$j_1$}
	 \psfrag{j_2}{$j_2$}
	 \psfrag{ln[eB^2]}{$Q$}
	 \psfrag{j_3=15/2}{}
	 \includegraphics[height=6cm]{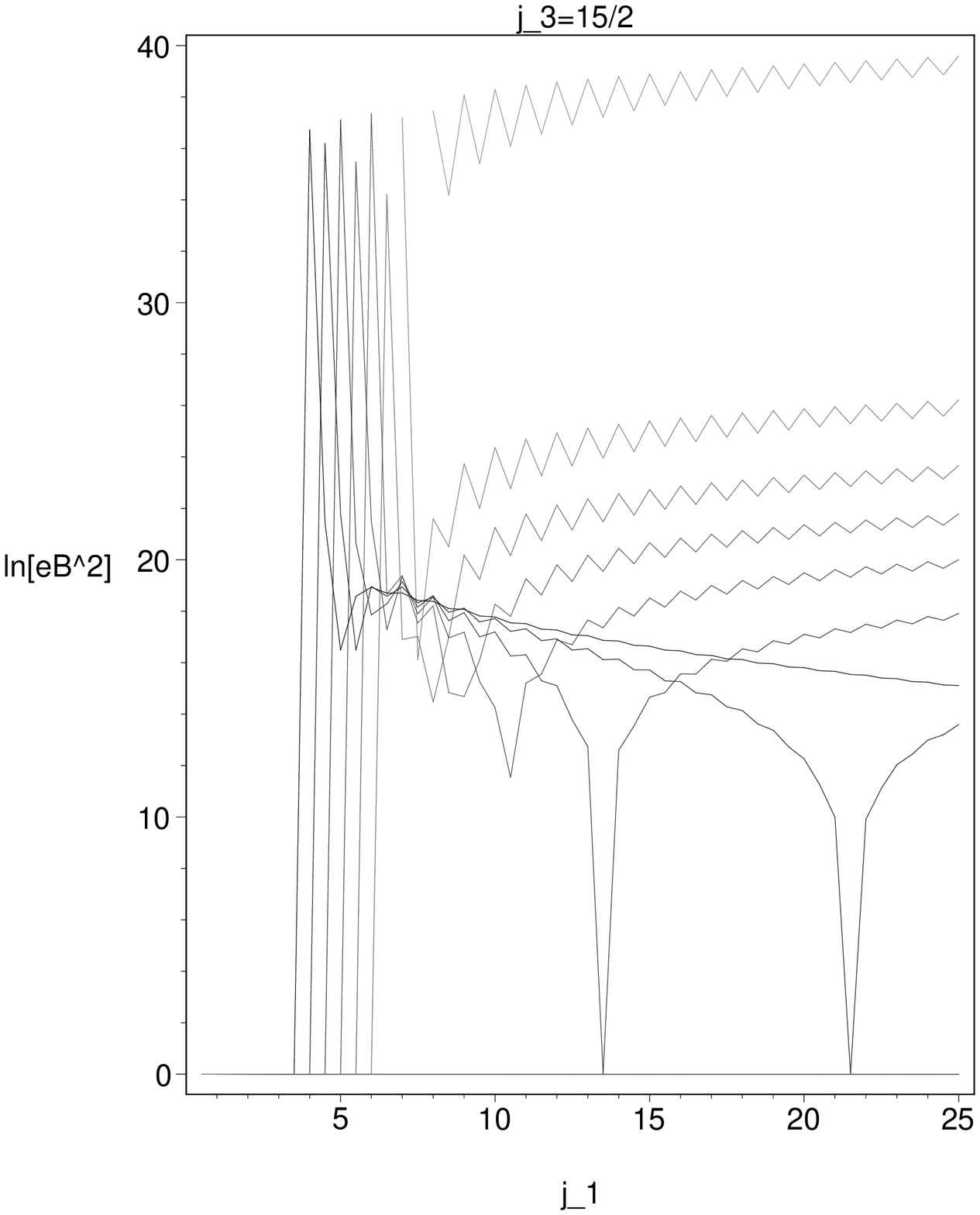}
          \caption{Plot for $j_3=\frac{15}{2}$, $j_2=j_1-l$. The different curves are (bottom to top):
	  $l=\frac{1}{2},\frac{3}{2},\ldots,\frac{15}{2}$} } 
\end{figure}
~\\

\pagebreak 
Due to the large kernel of the volume operator even for four -- and higher 
valent vertices \cite{17} a qualitatively similar behaviour will prevail 
in those cases as well. The unboundedness of the ``inverse scale factor''
operator in full LQG, even on homogeneous and isotropic states is herewith 
established: Let $\gamma_1,..,\gamma_n$ be a random distribution of
mutually disjoint but diffeomorphic 
graphs, say of the type of figure \ref{fig1} which is homogeneous and 
isotropic with respect to the comoving spatial metric of a FRW spacetime 
on a scale large compared to the Planck length within a compact 
region (the smaller that scale the larger $n$). The combined graph 
$\gamma:=\cup_{k=1}^n\; \gamma_k$ looks then homogeneous and 
isotropic. Consider the state of unit norm 
\be \label{3.5}
T_{\gamma,\lambda}:=\otimes_{k=1}^n \; T_{\gamma_k,\vec{j}} 
\ee
where we choose the same triple of spins for every graph and such that the 
numbers (\ref{3.4}) are equal to the same value $\lambda^2$ for every 
vertex of the $\gamma_k$. Then obviously all the norms (\ref{3.4}) with 
$T_{\gamma,\vec{j}}$ chosen as the homogeneous state (\ref{3.5}) equal 
$\lambda^2$ for every of the $2n$ vertices of $\gamma$. But $\lambda$ can 
be arbitrarily large. \\
\\
Let us compare this with the situation for the hydrogen atom: In both 
LQG and LQC
the spectrum of the inverse scale factor is purely discrete, so we should 
compare with the bound states of the hydrogen atom. Restricted to the 
subspace of bound states it is indeed the case that the Hamiltonian of the 
hydrogen atom is bounded from below and above and this is not changed 
when we switch from the finite number of degrees of freedom of QM to the 
infinite number of degrees of freedom of QED. However, the interplay 
between LQC and LQG is different: In LQC we have boundedness, but after 
switching 
on LQG we get unboundedness.

\section{Boundedness of the Inverse Scale Factor Expectation Values at the 
Initial Singularity}
\label{s4}

The result of the previous section sounds rather negative at first. 
However, it is not as we will
now explain. It just shows that {\bf the boundedness of the inverse scale 
factor cannot be a mechanism for avoidance of the local singularity 
in LQG.} 

Namely, what
we have shown is unboundedness on certain homogeneous and isotropic 
spin network states. However, that does not 
mean that the inverse scale factor is unbounded on all homogeneous 
and isotropic states. In fact, we should worry only about boundedness on 
those states which describe a collapsing universe which is homogeneous 
and isotropic on large scales. Indeed, in this section we will 
display the result of a calculation 
which indicates that 
the expectation value of the inverse scale factor with 
respect to (kinematical) coherent states \cite{20} peaked on any 
homogeneous and 
isotropic point in phase space (e.g. a FRW universe along the classical 
singular trajectory) {\bf is bounded from above in LQG even at the Big Bang.} 

The details of our calculation can be 
read in the companion paper \cite{16}. However, again we can explain the 
basic mechanism for this boundedness result in very intuitive terms:\\
Spin network states are, roughly speaking, eigenstates for the kinematical
geometric operators such as the triad operator out of which the inverse 
scale factor is assembled. They are thus maximally sharp for half of the 
degrees of freedom (the electric fields) but maximally unsharp for the 
other half (the conections) which are encoded in the 
holonomy operators. Coherent states intermediate in between those two 
uncertainties and in fact minimize the Heisenberg uncertainty bound as far 
as conjugate canonical pairs are concerned. This is achieved by 
superimposing basis states such as spin network states with carefully 
chosen coefficients. These coefficients depend on a point in the classical 
phase space, in our case a field configuration $(A,E)$ of connections and 
electric fields respectively. The coherent states are thus labelled by 
these points and they are peaked in phase space in the sense that the 
overlap function $|<\psi_{(A,E)},\psi_{(A',E')}>|^2$ vanishes 
exponentially fast unless $||A-A'||^2,\;||E-E'||^2\le s$ where $||.||$ is 
a suitable norm on the classical field space and $s$ is a parameter which 
encodes the Gaussian width of the wave packet or the volume of a cell in 
phase space (it is proportional to $\hbar$). The Gaussian decay of the 
overlap function implies that in terms of spin network states the state is 
of the form 
$\psi_{(A,E)}=\sum_j\; c_j(A,E)\; T_j$ where we have not displayed the 
graph label 
for simplicity and where the $c_j$ decay (symbolically) as $\exp(-s j^2)$. 
The label 
$j$ is a compound label for all participating spins and intertwiners.
Now while the norm of the inverse scale factor in the states $T_j$ is 
unbounded from above, that bound only diverges with a finite power as 
$j\to \infty$. This divergence therefore is completely swamped by the 
Gaussian decay of the exponential. This holds for rather generic values
of $A,E$ and in particular when we choose a homogeneous and isotropic 
point in the phase space. Hence the mechanism is similar as in statistical
physics: ``Energy'' (the argument of the Gaussian factor) wins over 
``entropy'' (the number of configurations leading to divergent 
eigenvalues). This a rather different mechanism which does not require
boundedness of the inverse scale factor.\\
\\
Let us go into more detail:\\  
Given a point $(A_0,E_0)$ in the classical phase space, our (not 
normalized) coherent states 
are of the general form 
\be \label{4.1}
\psi_{(A_0,E_0)}=\sum_\gamma\;\psi_{\gamma,(A_0,E_0)}
:=[e^{-\hat{{\rm\bf C}}/\hbar}\;\delta_{A'}]_{A'\mapsto A^\Cl(A_0,E_0)}
\ee
where the uncountable sum extends over all finite graphs and 
$\psi_{\gamma,(A_0,E_0)}$ is a countably infinite coherent superposition 
of 
spin network states over $\gamma$. The coefficients in that superposition 
are uniqely determined by the generator ${\rm\bf C}$ of the coherent 
states, also 
called the complexifier \cite{22a}, which is supposed to have the 
dimension of an action. In fact, as we can see from 
(\ref{4.1}), the coherent states are 
nothing else than the heat kernel (with respect to the operator 
$\hat{{\rm\bf C}}$) evolution of the $\delta$ distribution with support at 
$A'$ 
which is then analytically extended to $A^\Cl(A_0,E_0)$. The complexified 
connection $A^\Cl$ also is generated by ${\rm\bf C}$ via
\be \label{4.2}
A^\Cl:=\sum_{n=0}^\infty \;\frac{i^n}{n!}\;\{{\rm\bf C},A\}_{(n)} 
=A+i\;\{{\rm\bf C},A\}-\frac{1}{2}\;\{{\rm\bf C},\{{\rm\bf 
C},A\}\}-\frac{i}{6} \{{\rm\bf C},\{{\rm\bf C},\{{\rm\bf C},A\}\}\}+...
\ee
This may look unfamiliar, however, this is precisely the way that the 
unnormalized coherent states for the harmonic oscillator are constructed 
for which the 
complexifier reads ${\rm\bf C}=p^2/(2m\omega)$ and the classicality 
parameter 
is $s=\hbar/(m\omega)$ (the role of the connection is of 
course played by the configuration coordinate $q$). 

Ideally, according to the subsequent section, ${\rm\bf C}$ should be 
adapted 
to a given Hamiltonian, but as we 
said, here we will carry out a completely kinematical calculation to get a 
first insight. It is therefore appropriate to choose ${\rm\bf C}$ in such 
a way as 
to obtain the simplest possible states. One such choice involves the 
square of the area operator \cite{22a} which in turn is proportional to 
$E^2$ and thus will give rise to states very similar to the ones for the 
harmonic oscillator because in both cases the complexifiers are of the 
momentum squared type giving rise to Gaussian wave packets. Rather than 
defining ${\rm\bf C}$ we directly display 
the corresponding states which in that case are, roughly, of the direct 
product form
\be \label{4.3}
\psi_{\gamma,(A_0,E_0)}=\prod_{e\in E(\gamma)}\;\psi_{e,(A,E)}
\mbox{ where } 
\psi_{e,(A_0,E_0)}(A)=\sum_{2j=1}^\infty\;(2j+1)\;e^{-\frac{t(e)}{2}j(j+1)}
\;\mbox{Tr}(\pi_j(g_e(A,E)\;A(e)^{-1}))
\ee
Here $\pi_j$ is the spin $j$ irreducible representation of $SU(2)$. 
Notice that that the wave function $\psi_{e,(A_0,E_0)}(A)$ is a function 
of $A$ (or of the $SU(2)$ valued holonomies $A(e)$) which depends 
parametrically on $(A_0,E_0)$. The dimensionless parameter $t(e)$ depends 
on the details of ${\rm\bf C}$ and satisfies $t(e)>0,\;t(e\circ 
e')=t(e)+t(e'),\;
t(e^{-1})=t(e)$. Finally, 
$g_e(A_0,E_0)\approx \exp(i \tau_j E^j_0(S_e)/2)\;A_0(e)$ where 
$S_e$ is, roughly speaking, the face of a cell complex dual to $\gamma$ 
(i.e. each $S_e$ 
intersects $\gamma$ only in an interior point of $e$) which also depends 
on the details of ${\rm\bf C}$). 
Experts will see the representation theory of $SU(2)$ and the Peter\&Weyl
theorem at work in (\ref{4.3}). 

As explicitly shown in \cite{20}, expectation value computations with 
coherent states are qualitatively unaffected when we replace the group
$SU(2)$ by $U(1)^3$ for which however the analysis simplifies 
tremendously. The corresponding coherent states are then given by
\be \label{4.4}
\psi_{e,(A_0,E_0)}(A)=\sum_{|n_1|,|n_2|,|n_3|=1}^\infty\;
e^{-\frac{t(e)}{2}\sum_{j=1}^3 n_j^2}\;
\;\prod_{j=1}^3\; (g^j_e(A,E)\;A^j(e)^{-1})^{n_j}
\ee
where $A^j(e)=\exp(i\int_e A^j)$ and  
where 
\be \label{4.5}
g^j_e(A_0,E_0)=\exp(\int_{S_e} dS_a E^a_{0j}+i\int_e A_0^j)
\ee
To carry out concrete calculations we just have to insert concrete values 
for $A_0,E_0$ into those formulae. In cosmological applications we choose,
as in LQC (see section \ref{s0}), e.g. $A_{0a}^j=q\delta_a^j$ and 
$E^a_{0j}=p\delta_a^j$ where 
$a=\sqrt{|p|}$ is the scale factor. Thus in LQG, rather than performing a 
{\it cancellation} of the inhomogeneous degrees of freedom as in LQC, we 
perform a {\it suppression} of their fluctuations by using appropriate 
states.

What one actually does in the computations is not to use
the states $\psi_{(A_0,E_0)}$ but rather the ``cut -- off'' states
$\psi_{\gamma,(A_0,E_0)}$ for finite but sufficiently fine $\gamma$.
Roughly speaking, the analysis of \cite{20,22a} shows that the optimal 
graph, for given resolution scale $L$ of a measurement\footnote{Given a 
metric $g_0$ to be approximated, we want fluctuations in the length to be
much smaller than $L$ where $L$ is to be measured with respect to $g_0$. On 
the other hand, the Planck scale discreteness of LQG forces 
$\epsilon\gg L$.} 
should have an average edge length scale (with respect to the metric to be 
approximated) of the order of a geometric mean $\epsilon\approx L^k 
\ell_P^{1-k}$ where $0<k<1$ is a parameter that depends on the (partial)
observables to be approximated. 
The reason for working with the cut -- off states is that the 
$\psi_{(A_0,E_0)}$ are not normalizable 
on the kinematical Hilbert space due to the sum over the uncountable 
number of graphs. They merely serve as a generator for the cut -- off 
states which {\it are} normalizable.

What we have done then is to estimate from above the expectation value 
not only of the inverse scale factor in LQG for a general valence $M$ of the vertex 
for arbitrary $(A_0,E_0)$ but also the most general polynomials built out 
of generalized co -- triad operators coming from classical expressions of 
the form $(\{A,V^r\})^N$ which become upon quantization operators of the form 
$\big[\frac{\mb{i}}{\hbar\kappa}\big]^N\prod_{I=1}^N\hat{q}^j_{e_I}(v,r)=
\big[\frac{\mb{i}}{\hbar\kappa}\big]^N\prod_{I=1}^N\hat{h}^j_{e_I}\big[(\hat{h}^j_{e_I})^{-1},\hat{V}^r \big]$ where $r,N$ are fixed, positive, rational and natural numbers respectively which are determined by the concrete coupling between
matter and gravity \cite{10b}. Here $\hat{h}^j_e$ are the holonomies of the connection $A$ along edges of the graph, the classical volume expression $V$ turns into the volume operator $\hat{V}$ and Poisson brackets into commutators. A bound is given by 
\footnote{Neglecting all terms of order $t^\infty$.} 


\ba\label{finale obere Schranke fuer Erwartungswert}
    \lefteqn{\Big<~\Psi_{m,\gamma}^{(v)}(A)~\Big|~\prod_{k=1}^N 
\hat{q}^{j_k}_{I_k} (r)~\Big|~\Psi_{m,\gamma}^{(v)}(A)~ \Big>\le}
   \nonumber\\
   &\le&
(\ell_P)^{3rN}(9M)^N|Z|^\frac{rN}{2}\big[\frac{2\mb{A}}{T^2}\big]^N \;\;
     \Bigg[\bigg[\frac{3Mp}{4T}\bigg]^N+\sum\limits_{n=1}^{N}\frac{N!}{(N-n)!~n!} \bigg[\frac{3Mp}{4T}\bigg]^{N-n}
     \prod_{l=1}^n\bigg[\frac{3M+2(l-1)}{4}\bigg]\Bigg]~~~~~~~~~ 
\ea


where $M$ denotes the valence of the vertex, $N$ the degree of the polynomial, $Z$ is a constant numerical prefactor, depending on the Immirzi parameter and the regularization of the volume operator. 

Moreover 
\[\begin{array}{lllclc}
                                T:=\min\limits_I\{T_I\} &~~~~&\rightarrow~~
				&\mb{A}&:=1+\frac{p}{T}
				\\
				p:=\max\limits_{I,j}\{|p^j_I(m)|\}
				&&&          
                                
			     \end{array} \]

 with $T_I=\sqrt{t(e_I)}$, $t(e_I)$ is the classicality 
 parameter mentioned above and
 $p^j_I(m)\approx\frac{\epsilon}{L^3} \int_{S_{e_I}} dS_a E^a_j$.
 Here $m=(A_0,E_0)=(A,E)$ is the point in classical phase space the 
 coherent state is peaked on and $L,\;\epsilon$ respectively are the 
resolution scale and graph scale mentioned above. 

Specializing to homogeneous data  
$m=(A^j_{0a},E^a_{0j})=(\tilde{q}\delta^j_a,\tilde{p}\delta^a_j)$ gives 
 $p=\tilde{p}\cdot\max\limits_{I}C_I^{(construct)}$, with numerical 
constants
 $C_I^{(construct)}$ carrying information about the details of the 
coherent state. 


Taking the limit $a\to 0$ which causes $p=0, \mb{A}=1$ demonstrates that nothing dramatic happens
and we get the huge but finite bound 

\ba\label{obere Schranke fuer Erwartungswert im Urknall}
   \Big<~\Psi_{m,\gamma}^{(v)}(A)~\Big|~\prod_{k=1}^N \hat{q}^{j_k}_{I_k} (r)~\Big|~\Psi_{m,\gamma}^{(v)}(A)~ \Big>\bigg|_{a=0}   
   &\le&
(\ell_P)^{3rN}(9M)^N|Z|^\frac{rN}{2}\big[\frac{2}{T^2}\big]^N \;\;
      \prod_{l=1}^N\bigg[\frac{3M+2(l-1)}{4}\bigg]~~~~~~~~~ 
\ea

 For the case of the inverse scale factor we have to set $(r=1/2,\;N=6)$ 
and to multiply by $\frac{1}{(\hbar\kappa)^N}=\frac{1}{(\ell_P)^{12}}$. \\
\\
These results prove our claim: The expectation value (and thus also the 
norm, just double $N$) of the inverse scale factor is bounded with respect 
to kinematical semiclassical states even when the classical metric that 
labels the coherent state becomes maximally degenerate. As in LQC this is 
a strong indication 
for a drastic modification of the effective Friedmann equations near 
$a=0$ \cite{13}. This should be worked out in more detail in order to 
decide, along the scheme given in section \ref{s3}, whether the initial 
local singularity is avoided or not.\\
\\
Remarks:\\ 
1.\\
We have computed the expectation value of the 
analogon of the inverse 
scale factor with respect to kinematical coherent states whose peak 
follows the classcial (singular) trajectory. One should do the same with 
respect to the matter degrees of freedom. However, it is clear from the 
outset that this will not change our result: The classical Friedmann 
equations in the gauge outlined above are 
\be \label{4.6}
3H^2=8\pi G(\frac{\pi^2}{2a^6}+\frac{a^3 V}{a^3})+\Lambda-\frac{3k}{a^2},\;\;
\dot{\pi}=-a^3 V'
\ee
where $H=\frac{\dot{a}}{a}$ is the Hubble function and $\pi=a^3 
\dot{\phi}$
the conjugate momentum. Consider the case of a 
potential which depends polynomially on $\phi$ but not exponentially.
We want to solve (\ref{4.6}) close to $a=0$. We now make the 
self-consistent assumption (to be checked) that $a^3 V$ (and thus $a^3 
V'$) 
vanishes in the 
limit $a\to 0$. Hence $\pi$ becomes constant close to $a=0$ according to 
the 
second equation in (\ref{4.6}). Then we may neglect the 
potential ($V$), cosmological 
($\Lambda$) and curvature ($k$) term in the first equation as $a\to 0$
and thus we may solve $a(t)\propto t^{1/3}$ and $\phi(t)\propto \ln(t)$.
Hence indeed $a^3 \phi^n\propto t [\ln(t)]^n$ vanishes at $t=0$ and our 
assumption was correct. It follows that the expectation value with respect 
to coherent states, taking into account both matter and geometry, of the 
scalar contribution to the Hamiltonian constraint 
(\ref{0.7}), whose homogeneous reduction is 
$\pi^2/(2a^3)+a^3 V$, is indeed non-singular even when the coherent states 
are evaluated along the {\it exact} classical trajectory, if and only if 
the expectation value of the analogon in (\ref{3.3}) of $1/a^3$ is bounded 
from 
above. This is precisly what we verified. \\
2.\\
We have shown unboundedness of the inverse scale factor in LQG (with gauge 
group $SU(2)$) and did expectation value calculations therefof for an 
approximation to LQG using $U(1)^3$ instead of $SU(2)$. It is maybe 
worthwhile to mention that the inverse scale factor for that approximation 
is also unbounded on zero volume eigenstates although one has to use 
graphs of valence at least five here which is due to the detailed 
difference between the non -- Abelean group $SU(2)$ and the Abelean group
$U(1)^3$.\\
3.\\
Notice that the bound in formula (\ref{finale obere Schranke fuer 
Erwartungswert}) is completely general. It does not depend on the value of 
the classical connection $A$ that labels the coherent state, it only 
depends on the electric field $E$. Setting $E^a_j=0$ as appropriate 
for the isotropic singularity gives the  
upper bound (\ref{obere Schranke fuer Erwartungswert im Urknall}). 
However, we may consider any other homogeneous model, for instance 
inhomogeneous ones such as Bianchi I (Kasner), Bianchi IX or Mixmaster 
Models (BKL scenario). In these cases the different components of the 
three metric vanish with different rates as we approch the singularity
(or do not vanish at all or might even diverge, at least in the absence 
of matter). As we can see from the above general estimate, the important 
question is whether the quantity 
$p$ defined above, which is proportional to $E^a_j=\sqrt{\det(q)} e^a_j$,
remains finite in all those cases. For instance, in the Kasner 
models with 
line element $ds^2=-dt^2+\sum_{a=1}^3 t^{2p_a} (dx^a)^2$ where the 
real parameters $p_a$ are subject to $\sum_a p_a=\sum_a p_a^2=1$ we find 
in the diagonal gauge $E^a_j=\delta^a_j t^{1-p_a}$ which stays bounded 
for $t\to 0$ because $p_a\le 1$. In particular, contracting 
directions (with increasing $t$) are allowed. This covers all 
Kasner models and shows that the expectation value calculation 
performed here also hints at local singularity 
avoidance for at least some class of anisotropic models. 
Whether this holds for all homogeneous Bianchi models and also after
matter is coupled is a task left to the reader.

\section{Scheme for Determining the Presence or Absence of the Initial 
Singularity}
\label{s3}

The computation of the previous section is very encouraging.
Unfortunately, it does not yet prove that in LQG the 
Big Bang singularity or any other classical spacetime curvature 
singularity is 
quantum mechanically absent. As we have explained at length in the 
introduction, fundamentally, singularity avoidance 
must be formulated in terms of physical states\footnote{The singularity 
effects A and B for the global singularity have indeed been discussed for 
solutions of the Wheeler DeWitt equation in LQC. However, that does not 
grant
that a sufficient number of them are normalizable or have non vanishing 
norm in the physical inner product. Once again, formal solutions are not 
automatically physical states.}
and physical operators 
which has not been done so far even in LQC\footnote{In terms of the 
the partial and complete observables discussed below, there is in general
no relation between the spectra of the complete and partial observables
\cite{20a}. Hence kinematical boundedness does not imply physical
boundedness as far as the local singularity is concerned.}
because the volume, the inverse scale factor and the value of the inflaton 
field etc. are not gauge invariant quantities (they do not commute with 
e.g. the Hamiltonian constraint). 

Fundamentally, what one would need to do 
in order to establish a singularity avoidance result is something along 
the following lines\footnote{This is just a sketch of ideas, not a 
``paradigm''.}: 
\begin{itemize}
\item[1.] Ideally one should have sufficiently explicit knowledge about 
the 
physical Hilbert space ${\cal H}_{Phys}$ and physical states. An avenue 
towards this goal is the Master Constraint Programme \cite{11}. Notice 
that physical states automatically solve the Quantum Einstein Equations.
That means that the quantum gauge evolution \cite{10b}, defined 
similarly as in \cite{12} as the coefficients of the physical states 
when expanded into (spatially diffeomorphism invariant) spin network 
functions, is non -- singular by definition thanks to the fact that 
operators
like the inverse scale factor are densely defined in LQG. 
Moreover, in LQG there are 
physical states which are composed of spin network states 
that describe a sign flip with respect to the expectation value of the 
triad orientation sign operator \cite{20b} and thus may describe quantum 
geometries which include a Pre Big Bang in the sense of \cite{12}.
As we have already outlined in the introduction, the global 
singularity is avoided if and only if there are sufficiently many 
semiclassical of these physical states. 
\item[2.] (In)famously, canonical quantum gravity, especially in the
cosmological context, is a theory without a canonical Hamiltonian. 
However, one can construct physical Hamiltonians using the relational 
point of view \cite{15,15a}. These are Dirac observables by construction.
They are obtained by i) using non -- gauge invariant quantities $P,T$
where $P$ is called a partial observable and $T$ a collection 
of\footnote{In general one needs as many clock variables as there are 
constraints.} clock variables,
ii) computing the gauge flows (Hamiltonian evolution with respect 
to the Hamiltonian constraints\footnote{There are also as many unphysical 
time parameters $t$ as there are constraints. In GR they become 
functions and are called lapse and shift.}) 
$t\mapsto 
\alpha_t(P),\;\;\alpha_t(T)$, 
iii) choosing parameters $\tau$ and solving the equations 
$\alpha_t(T)=\tau$ for $t$ and iv) inserting those values of $t$ 
into 
$\alpha_t(P)$. The resulting object $P_T(\tau)$ has the physical 
interpretation of giving the value of $P$ at the moment (gauge parameter 
$t$) when $T$ has attained the value $\tau$. The function 
$P_T(\tau)$ is called the corresponding complete
or Dirac observable. The physical Hamiltonians $H_T(\tau)$ are in general
explicitly dependent on the physical time $\tau$ and are defined by 
$\{H_T(\tau),P_T(\tau)\}:=d/d\tau P_T(\tau)$ for all $P$. Of course there 
are many 
possible choices for $H_T$ depending on the choice of the ``clocks'' $T$
(in GR we need an infinite number of those). In cosmology it would be 
nice to use the scale factor (total volume of space) as a clock. 
If a corresponding positive (or at least semibounded) and physical time 
independent Hamiltonian exists and can be 
defined as a self -- adjoint operator on the physical Hilbert space then
we may use it to define evolution on the physical Hilbert space. 
\item[3.] If a physical Hamitonian is given, then we will need coherent 
states in ${\cal H}_{Phys}$ which are preserved by $\hat{H}_T$ for 
sufficiently long physical time. A general 
idea for how to do that has been given by Klauder \cite{20} at least for 
systems with a finite number of degrees of freedom.
\item[4.] We are now in the position to address the issue about the 
local singularity: Given a physical coherent state $\psi$ peaked, on 
scales 
large 
compared to the Planck scale, corresponding to $\tau=\tau_0\gg \tau_P$, on 
classical 
singular initial data, consider 
the map 
$\tau\mapsto <\psi,\widehat{P_T(\tau-\tau_0)}\psi>_{Phys}=
<\psi(\tau-\tau_0),\widehat{P_T(\tau_0)}\psi(\tau-\tau_0)>_{Phys}$
with $\psi(\tau):=e^{i\tau\hat{H}_T}\psi$. Let $P$ be any classically 
singular quantity, say the 
matter energy density or a curvature scalar density (suitably smeared). 
Then the local 
singularity, with respect to $P$, is absent provided that
$<\psi,\widehat{P_T(-\tau_0)}\psi>_{Phys}$ is finite (assuming that 
$\tau=0$ corresponds to vanishing classical volume which will be the 
case if we choose $V$ as one of the clock variables). 
\end{itemize} 
Notice that this ambitious scheme has not been followed even in the 
simplified context of LQC where, however, all the steps could presumably 
be carried out sufficiently explicitly, see \cite{22b} for first steps 
into that direction. In full LQG these steps cannot be 
carried out exactly because the theory is too complicated. Thus, in 
order to make progress one must use suitable approximation schemes. A 
possible starting point could be the following:
\begin{itemize}
\item[1'.] Using kinematical coherent states $\psi_{A,E}$ peaked on the 
constraint 
surface of the phase space we arrive at a subspace of the kinematical 
Hilbert space ${\cal H}_{Kin}$ with the property that the constraints 
are satisfied approximately in the sense of expectation values, that is
$<\psi,\hat{C}\psi>\approx 0$. Denote that subspace 
by ${\cal H}'_{Phys}$ \footnote{Alternatively, find an approximation 
method for the Master Constraint 
Programme (also using kinematical coherent states and path integral 
methods, see \cite{11}).}. In systems where one can compare predictions by 
the exact ${\cal H}_{Phys}$ with those by ${\cal H}'_{Phys}$ it is often 
the case that the qualitative answers do not differ much. See e.g. 
\cite{22}. Since there are as many semiclassical kinematical states as
classical initial data (classical spacetimes), global singularities cannot 
be 
determined within this approximate scheme which uses the kinematical 
Hilbert space.
\item[2'.] The formulas for $P_T(\tau)$ and $H_T$ are infinite power 
expansions in the ``parameters'' $\tau-T$ \cite{15a}. Hence, close to the 
singularity,
i.e. when $\tau\approx T=0$ (this has to be understood ultimatively in 
terms 
of operators; we assume that the $T,\tau$ are chosen such that 
$T=\tau=0$ corresponds to the singularity) we may 
approximate those formulas by the first few terms and 
get approximate expressions $\hat{P}'_T(\tau),\;\hat{H}'_T$ that one can 
hope to handle. These approximations should be appropriate in order to 
answer qualitatively questions about the very early universe.
\item[3'.] The coherent states of item 1' should ideally be chosen as to 
be preserved by the approximate Hamiltonian determined in step 2'. If that 
is too difficult, peak the kinematical coherent states of step 1' on
classically singular initial data and then quantum evolve them with the 
(approximate) Hamiltonian $\hat{H}'_T$, that is 
$\tau\mapsto \exp(i\tau \hat{H}'_T)\psi_{A,E}$.
\item[4'.] This step is identical to step 4 above with the obvious 
replacements. 
\end{itemize}
Notice that one could consider an even easier scheme: Just take any set of 
kinematical coherent states not generated by any (approximate) physical 
Hamiltonian and consider the unphysical time evolution 
$t\mapsto \psi_{A_0,E_0}(t):=\psi_{A(t),E(t)}$ where $A(t),E(t)$ is 
a one 
parameter classical gauge 
evolution of the initial data under the flow of the 
corresponding Hamiltonian 
constraint (not of any physical Hamiltonian)\footnote{This 
corresponds to a specific choice of lapse and shift.} with initial
condition $A(t_0)=A_0,\;E(t_0)=E_0$ and $t_0\gg t_P$. 
Then take any quantity $P$ 
which is not necessarily a (approximate) Dirac observable and study
$<\psi_{A(t),E(t)},\hat{P}\psi_{A(t),E(t)}>_{Kin}$ in the limit $t\to 0$
corresponding to the classical singularity (in our case $A(t)$ diverges 
while $E(t)$ vanishes at $t=0$). This is essentially what we 
did in the previous section where $P$ played the role of the 
inverse scale factor and the coherent states we used are those of 
\cite{20}. Wile in this case we obtained a boundedness result, one may not 
draw any conclusions from that as the following example reveals:\\
Consider the Hamiltonian of the hydrogen atom. Let 
$\psi_{\vec{q},\vec{p}}$ be the coherent states of the harmonic oscillator 
peaked at the point $(\vec{q},\vec{p})$ in phase space. Let 
$\vec{q}(t),\vec{p}(t)$ be the trajectory of a classical electron falling
radially into the proton from a distance large compared to the atomic 
radius where $t=0$ is the point of time when the 
electron reaches the center. Then it is easy to see that 
$<\psi_{\vec{q}(t),\vec{p}(t)},\hat{H} \psi_{\vec{q}(t),\vec{p}(t)}>$ 
diverges as $t\to 0$. This is because the electron becomes infinitely fast 
at the center and the expectation value of the kinetic term diverges
(while the expectation value of the potential energy stays finite). In 
contrast, the expectation 
value
$<\psi_{\vec{q},\vec{p}}(t),\hat{H} \psi_{\vec{q},\vec{p}}(t)>
=<\psi_{\vec{q},\vec{p}},\hat{H} \psi_{\vec{q},\vec{p}}>$ is finite where 
$\psi_{\vec{q},\vec{p}}(t):=\exp(it\hat{H})\psi_{\vec{q},\vec{p}}$
is the quantum evolved state\footnote{Remarkably, the states 
$\psi_{\vec{q},\vec{p}}(t)$ and $\psi_{\vec{q}(t),\vec{p}(t)}$ precisely 
coincide if and only if the underlying dynamics is that of a finite or 
infinite sum of uncoupled harmonic oscillators.}. Thus, while 
$\psi_{\vec{q}(t),\vec{p}(t)}$ is nicely peaked on the classical 
trajectory, it does not correspond to the true quantum mechanical time 
evolution. On the other hand, the true quantum mechanical time evolution 
$\psi_{\vec{q},\vec{p}}(t)$ usually does not stay peaked on the classical 
trajectory for sufficiently long time in order to qualify as a 
semiclassical state representing the classical evolution far away from 
the deep quantum regime closely enough, unless the coherent state is 
actually adapted to the Hamiltonian $\hat{H}$. \\
\\
Summarizing, ideally
the coherent states must be chosen according to the Hamiltonian in 
question in order to arrive at reliable predictions. If that is too hard, 
then at least the kinematical coherent state should be evolved quantum 
mechanically but only as long as it remains close to the classical 
trajectory for scales above which the classical theory is reliable. In 
contrast, the classically evolved 
kinematical semiclassical state not constructed according to the 
Hamiltonian is inappropriate in general which is why the calculation of 
section \ref{s4} is promising but inconclusive.

\section{Conclusions}
\label{s5}

Concluding, the issue of the presence or absence of the initial
singularity in full LQG remains open. The results of LQC are 
important, they are very 
promising, provide important consistency checks for LQG and produce new 
insights and ideas about LQG. However, as we 
hope to have demonstrated, they are so far inconclusive for LQG. 
Therefore sentences such as ``LQC is the cosmological 
sector of LQG'' or ``in LQG the Big Bang is absent'' \cite{23}
are misleading for the non expert and should be used with due care.

As our detailed analysis 
reveals, one must pay attention to the QFT aspects of the full theory 
and cannot rely on the analysis of a quantum mechanical toy model.
This is, unfortunately, different from the case of the hydrogen atom 
where experiments confirmed from the outset that the quantum mechanical 
toy model was close to the truth and that the more accurate QED should 
only give corrections. 
It may still be that LQC is qualitatively right but if that 
is the case then the mechanism for singularity resolution is very 
different in full LQG and much more elaborate, if LQG is a valid candidate 
for quantum gravity at all. 

In order to resolve the issue about the initial singularity and other 
cosmological questions in full LQG we have proposed an ideal and 
approximate programme, highlighting the importance of having control over 
the physical Hilbert space and the physical observables. In a zeroth order 
step of that programme we found 
hints pointing at a resolution of the local initial singularity in LQG by 
a very 
different mechanism. But it is necessary to carry out all the steps before 
definite conclusions can be reached.
Since this programme was not even completed in LQC, again 
the LQC model could be a good starting point to see where the ideal path
leads us in a solvable situation. As far as the full theory is 
concerned we will only be able to follow the approximate path and we must 
learn how to control the corresponding approximation errors. \\
\\
In summary, the investigation of the comological or black hole sector of 
LQG, about 
which the LQG inspired LQC model has taught us important first lessons,  
is an ambitious and  
fascinating research field which will require the joint effort of 
both cosmologists and quantum geometers in the close future. \\
\\ \\ 
\\
{\large Acknowledgements}\\
\\
It is our pleasure to thank Martin Bojowald for many clarifying comments 
and a careful reading of the manuscript.
J.B. thanks the Gottlieb Daimler-- and Karl Benz--Foundation and the DAAD 
(German Academic 
Exchange Service) for financial support. This work was supported in part
by a grant from NSERC of Canada to the Perimeter Institute for Theoretical 
Physics.

\end{document}